\documentclass[a4paper,fleqn,usenatbib,useAMS]{mnras}

\usepackage{graphicx}	
\usepackage{amsmath}	
\usepackage{amssymb}	
\usepackage{multicol}        
\usepackage{bm}		
\usepackage{pdflscape}	

\usepackage[flushleft]{threeparttable}

\newcommand{\gsim}{\;\lower.6ex\hbox{$\sim$}\kern-7.75pt\raise.65ex\hbox{$>$}\;}
\newcommand{\lsim}{\;\lower.6ex\hbox{$\sim$}\kern-7.75pt\raise.65ex\hbox{$<$}\;}

\title[SSH I. Survey Description]{The Smallest Scale of Hierarchy Survey (SSH). I. Survey Description. }

\author[F. Annibali et al.]{
F. Annibali,$^{1}$\thanks{E-mail: francesca.annibali@inaf.it}
G. Beccari,$^{2}$
M. Bellazzini,$^{1}$
M. Tosi,$^{1}$
F. Cusano,$^{1}$
D. Paris,$^{5}$
\newauthor
M. Cignoni,$^{3}$
L. Ciotti,$^{4}$
C. Nipoti$^{4}$
E. Sacchi.$^{6}$
\\
$^{1}$INAF - Osservatorio di Astrofisica e Scienza dello Spazio, Via Piero Gobetti, 93/3, I-40129 - Bologna, Italy\\
$^{2}$ESO, Karl-Schwarzschild Strasse 2, D-80 Garching, Germany\\
$^{3}$Dipartimento di Fisica, Universit\`a di Pisa, Largo Bruno Pontecorvo 3, I-56127 Pisa, Italy\\
$^{4}$Dipartimento di Fisica e Astronomia, Universit\`a di Bologna, via Piero Gobetti 93/2, I-40129 - Bologna, Italy\\
$^{5}$INAF-Osservatorio Astronomico di Roma, Via Frascati 33, I-00078 Monte Porzio, Italy\\
$^{6}$Space Telescope Science Institute, 3700 San Martin Drive, Baltimore, MD 21218, USA\\
}

\date{Accepted XXX. Received YYY; in original form ZZZ}

\pubyear{2018}

\begin{document}
\label{firstpage}
\pagerange{\pageref{firstpage}--\pageref{lastpage}}
\maketitle

\begin{abstract}
The Smallest Scale of Hierarchy (SSH) survey is an ongoing  strategic large program at the Large Binocular Telescope, aimed at the detection of faint stellar streams and satellites around 45  late-type dwarf galaxies located in the Local Universe within $\simeq$10 Mpc. SSH exploits the wide-field, deep photometry provided by the Large Binocular Cameras in the two wide filters $g$ and $r$.
This paper describes the survey, its goals, and the observational and data reduction strategies. 
We present preliminary scientific results for five representative cases (UGC~12613, NGC~2366, UGC~685, NGC~5477 and UGC~4426) covering the whole distance range spanned by the SSH targets. We reach a surface brightness limit as faint as  $\mu(r)\sim$ 31 mag arcsec$^{-2}$ both for targets closer than 4$-$5 Mpc, which are resolved into individual stars, and for more 
distant targets through the diffuse light.   
Our analysis reveals the presence of extended low surface brightness stellar envelopes around the dwarfs, reaching farther out than what traced by the integrated light, and as far out as, or even beyond, the observed H~I disk. Stellar streams, arcs, and peculiar features are detected in some cases, indicating possible perturbation, accretion, or merging events.  We also report on the discovery of an extreme case of Ultra Diffuse Galaxy ($\mu_g(0)=27.9$~mag/arcsec$^2$) in the background of one of our targets, to illustrate the power of the survey in revealing extremely low surface brightness systems.

\end{abstract}

\begin{keywords}
galaxies: dwarf -- galaxies: formation -- galaxies: interactions -- galaxies: irregular -- galaxies: individuals: UGC~12613, NGC~2366, UGC~685, NGC~5477, UGC~4426  -- galaxies: stellar content.  
\end{keywords}


\section{Introduction} \label{intro}

In the $\Lambda$ Cold Dark Matter ($\Lambda$CDM) cosmological scenario \citep{Peebles82}, galaxies are assembled over time through the accretion of smaller systems \citep{White78}. In fact, there is ample observational evidence of interaction of satellites with the massive galaxies they are orbiting around, such as the Milky Way, Andromeda, spirals and giant ellipticals in the Local Volume \citep[e.g.,][]{Belokurov06,Ibata01,McConnachie09,Martinez10,pisces}. 
Actually, numerical models predict that isolated dark halos and sub-halos contain substructures down to the lowest resolution limit of the simulations  \citep{Diemand08,Wheeler15}. Hence, not only massive galaxies, but also low-mass ones are predicted to show observational evidence of satellite accretions. A few recent discoveries have indeed revealed the existence of satellites around dwarf galaxies, but their sample is still limited \citep[e.g.,][]{Rich12,Martinez12,Sand15,Belokurov16,Amorisco14,Annibali16,Makarova18}. Systematic searches for dwarf galaxy satellites are therefore important to observationally test the self-similarity of the hierarchical formation process at all scales. 

So far, the most detailed searches for dwarfs' companions have concentrated on  the satellites of the Milky Way (MW) and, to a lesser extent, of M31. Since these are  predominantly of early morphological type, because of the morphology-distance relation \citep{Dressler80,Binggeli87}, the vast majority of late-type dwarfs in the Local Universe have not been searched for small companions with deep wide-field imaging observations yet. 
The notable exceptions are of course the Magellanic Clouds, for which the Gaia satellite has recently allowed to measure precise proper motions and to confirm the existence of at least 4 sub-satellites \citep{Kallivayalil19}. However, late-type dwarfs are particularly interesting to search for the presence of stellar streams or companions, because of the possible connection between accretion events and star formation activity. Recent studies have shown that disturbed H~I kinematics, H~I  companions, and filamentary H~I structures  are more common in starburst dwarfs  than in typical star-forming irregulars  \citep{Lelli14}. 
Understanding the origin of starbursts in dwarf galaxies is important for several reasons: 
these systems are currently the primary candidate source for cosmic reionization at $z>6$ \citep{Atek14}  and, due to their shallow potential well, they are the best systems to study galactic winds and feedback from massive stars.

There are two ongoing surveys that can significantly improve the census of dwarf galaxy companions in the Local Volume \citep[but see also][for a survey of dwarf galaxy pairs at $0.005<z<0.07$]{Stierwalt15}. One is the Solitary Local dwarfs survey (Solo), a wide-field photometric study targeting all the 42 isolated dwarf galaxies within 3 Mpc of the MW and beyond the nominal virial radius of MW and M31 \citep{Higgs16}. Solo is based on multiband imaging from the Canada-France-Hawaii Telescope/MegaCam for northern targets, and the Magellan/Megacam for southern targets. The survey includes both irregular and spheroidal dwarfs, with a prevalence of late-type systems over early-type ones.
The other survey is the Magellanic Analog Dwarf Companions And Stellar Halos (MADCASH), designed to use resolved stars to map the virial volumes of dwarf spiral galaxies within $\lesssim$ 4 Mpc and with stellar masses of $1\mbox{--}7\times {10}^{9}\,{M}_{\odot }$ \citep{Carlin16}. MADCASH is based on Subaru/Hyper Suprime-Cam wide-field imaging.

In this paper we introduce a new survey aimed precisely at searching satellites of (and signs of accretion events around) late-type dwarfs at the smallest mass scale: the Smallest Scale of Hierarchy (SSH) survey. 
SSH is currently in execution at the Large Binocular Telescope (LBT). LBT, with the resolving power of its two primary mirrors of 8.4 m diameter each (i.e. a combined collecting area corresponding to that of a single 11.9 m mirror) and the large field of view (23$^{\prime} \times 23^{\prime}$) of its Large Binocular Cameras (LBC) offers the optimal observational set-up to study the outskirts of late-type dwarfs in the Local Universe. With these data we plan to characterise the frequency and properties  of streams and substructures around late-type dwarfs as a function of galaxy mass and environment, and to study the impact of accretion/merging phenomena on the system's star formation activity. 

SSH will observe 45 late-type dwarfs at distances between $\sim$1 and $\sim$10 Mpc in the $g$ and $r$ photometric bands (sampling 7$-$70 kpc auround the target), thus significantly extending the volume covered by the other two surveys. Only a few SSH targets, with distances $\lesssim$3 Mpc, overlap with Solo: UGC~8091, UGC~9240, UGC~12613, UGC~6817, UGCA~276, NGC~4163 and UGC~7577.  Moreover, the SSH dwarfs have luminosities/masses in most cases lower than those sampled by MADCASH, and there is only one galaxy, namely NGC~4214, in common between the two samples. In conclusion, there is only marginal overlap of our survey with Solo or MADCASH. 

Before embarking in such a long-term project, we tested its feasibility with a pilot study searching for streams or companions around DDO~68, a small star-forming dwarf \citep[$\sim10^8M_{\odot}$ in stars,][]{Sacchi16}, extremely metal-poor \citep[$\sim$1/40 $Z_{\odot}$,][]{Izotov09,Annibali19a},  located in the huge Lynx-Cancer void at $\sim$12.7 Mpc  \citep{Sacchi16}. 
Our LBT study revealed a few
previously undetected, faint ($\mu_r\sim29$ mag arcsec$^{-2}$) stellar substructures around the main body \citep{Annibali16} of DDO~68, one of them later studied in more detail with HST photometry and confirmed to be actually connected to DDO~68 \citep{Annibali19b}. 
This photometry, combined with N-body simulations, showed that DDO~68 is experiencing a merging with a ten times lighter companion, and is also accreting an even smaller system, 100 times lighter. This provides indisputable observational evidence of the hierarchical merging process in action at very low galactic mass scales. It also demonstrates the excellent performances of LBT in the search of substructures around dwarfs.

As suggested by the case of DDO~68, dwarfs at the outskirts of groups, or in isolation, are likely the best candidates where to search for the presence of "sub-satellites''. Cosmological  simulations show that,
once a group of dwarfs with their satellites falls into a massive host, like the MW, it is eventually disassembled by tidal forces wiping out evidence of coherent structure \citep{Deason15}. Therefore it is natural to expect satellites of satellites residing in lower density environments to survive longer than those residing close to massive hosts \citep[e.g.,][]{Bellazzini13}. 

The paper is structured as follows. 
Section~\ref{section_objectives} gives a general overview of the survey, its goals and observational strategy. The sample of galaxies is described in Section~\ref{section_sample}, the image acquisition in Section~\ref{section_observations}, and the image 
reduction in Section~\ref{section_reduction}.  Section~\ref{section_photometry} describes the photometric reduction, while we present  color-magnitude diagrams (CMDs) and star count density maps for five representative galaxies in Sections~\ref{section_cmd}  
and ~\ref{section_starmap}, respectively. The derivation of surface brightness profiles is described in Section~\ref{section_profiles}. We discuss our results in Section~\ref{section_discussion} and present a summary in Section~\ref{section_conclusions}.

\section{The LBT SSH Survey} \label{section_objectives}

SSH is an Italian  {\it strategic\footnote{{\it Strategic} programs are long-term programs with the highest priority meant to increase the impact of LBT in the astronomical community at large by leading to significant advances in key scientific topic, highlighting the capabilities of the LBT unique instruments.}} long-term program (PI F. Annibali) in execution at LBT since the end of 2016. 
As mentioned above, the main goals of SSH are: i) characterizing the frequency and properties  of streams and substructures around late-type dwarfs as a function of galaxy mass and environment, and ii) linking the accretion/merging events with the galaxy star formation history (SFH) derived from deep CMDs obtained with Hubble Space Telescope (HST) photometry. These two long-term goals require the study of a conspicuous sample in order to provide statistically significant results, and thus will be reached only after completion of our program. 
However, as with our pilot study of DDO~68, successful detection of sub-structures in individual targets may provide by itself very interesting results.

The target galaxies, located up to a distance of $\sim$10 Mpc, cover a wide range of density environments, from very isolated galaxies to group members, and a wide range in luminosity, from $\sim$ twice as luminous as the LMC, to 5 magnitude dimmer. 

Our strategy is to search for faint substructures on $g$- and $r$- images obtained with LBC 1~h binocular exposures, using stellar counts for targets closer than $\sim$ 4$-$5 Mpc  (the exact distance depending on the seeing and on the galaxy  crowding), reaching a surface brightness limit as faint as  $\mu_r\sim31$ mag arcsec$^{-2}$  
\citep[see Section~\ref{section_profiles} of this paper and the study of Centaurus A by][]{pisces}, and relying on the integrated light for more distant targets, reaching a 
similar surface brightness limit (see Section~\ref{section_udg}). Observations in two bands are necessary to construct, for the nearest galaxies,  CMDs that can allow to separate red giant branch (RGB) stars from background contaminants. For all the galaxies, $g$ and $r$ photometry will be used to constrain the age, the metallicity, and the stellar mass of the detected substructures. Candidate companions or substructures detected around targets more distant than 4$-$5 Mpc will need further  follow-up observations (spectroscopy and/or high spatial 
resolution imaging) to constrain their distance and confirm their physical association with the host galaxies \citep[see e.g.][for the case of DDO~68]{Annibali16,Annibali19b}. 

Finally, we will run N-body simulations to reconstruct the system's interaction history. The simulations will  constrain the mass  and structural and orbital parameters of the accreting satellites, as well as the timescales of the interaction.

SSH will also allow us to investigate the hypothesis that strong starbursts in dwarf galaxies are triggered by interaction/accretion events. For instance, in the case of DDO~68, a significant increase in the star formation activity occurred $\sim$100 Myr ago, compatible with the encounter timescale predicted by our  
N-body simulations.  Is there always a temporal correspondence between periods of enhanced star formation activity and a close encounter, as derived from N-body simulations? To try to  answer this question, we will homogeneously re-analyse archival HST data of our sample targets and use our code \citep{Cignoni15} to infer the SFH from their CMDs. Then we will look for possible relationships between the star formation activity and the galaxy accretion history.

\section{The Sample} \label{section_sample} 

 \begin{table*}
  \caption{The SSH Sample.}
  \label{sample_tab}
  \begin{tabular}{llccccccccccc}
\hline
Name &  Other Name &  R.A.(J2000) & Dec.(J2000) & $a_{26 }$ & Distance & $a_{26 }$ & $M_B$ & T & $A_V$ & $\theta_1$ & MD & $\theta_5$      \\
& & [hh:mm:ss]  & [dd:mm:ss] &  [arcmin]  & [Mpc] &  [kpc] & [mag] & & [mag] & & & \\
  \hline \hline 
UGC~685     &                         &  01:07:22.4	& +16:41:02   &1.2	& 4.70   &  1.6   & -14.3  &   9    &    0.157   & -1.4 & NGC~0253  &   -1.1   \\
UGC~1249   &   IC~1727        &   01:47:29.9    & +27:20:00   & 8.0   & 7.45   & 17.3  & -18.1  &   8    &    0.217   &  4.0 & NGC~0672   &  4.0 \\
UGC~1281   &                         &    01:49:32.0    & +32:35:23   & 4.7   & 5.13   &  7.0   & -16.7  &   7    &    0.130   &  -1.1 & NGC~0784  &   -0.9\\
KK~16          & AGC~111977   &    01:55:20.6    & +27:57:15   &1.0    & 4.7     &  1.4    & -12.1 &  10   &    0.187   & -1.5  & NGC~0784  &  -1.1 \\ 
KK~17          & AGC~111164   &   02:00:09.9    & +28:49:57   & 1.0   & 4.7     &  1.4    & -11.2  &  10  &    0.146    & 1.1  & NGC~0784   &   1.1\\
NGC~784     & UGC~1501      &   02:01:17.0	& +28:50:15   & 6.6	& 4.94   &   9.5   & -16.1  &   7   &    0.162    & -1.1 & NGC~0784  &  -0.9  \\
KKH~37	    &                         &   06:47:45.8	& +80:07:26   & 1.2	& 3.39   &  1.2   & -11.6   &  10   &   0.204    & 0.0 & M~81           & 0.3  \\
UGC~3755   &                         &   07:13:51.6    & +10:31:19   & 1.8   & 6.66   &  3.5   & -15.9   &  10   &   0.242   & -2.2 & NGC~2683 &    -1.7\\  
NGC~2366   & UGC~3851     &  07:28:54.6	& +69:12:57   & 7.3	& 3.19   &   6.8   & -16.2  &  10   &   0.100   & 1.0  & NGC~2403  &    1.1  \\
UGC~3974   & DDO~47         &     07:41:55.4    & +16:48:09   & 5.0   & 8.02   & 11.7   & -16.0  &    8   &    0.091  & 1.0 & KK~65       &  1.0  \\
KK~65           &                      &    07:42:31.2    & +16:33:40   & 1.1    & 8.02   &  2.6   & -14.3  &   10  &   0.085    & 1.6 & UGC~3974      & 1.6 \\
UGC~4305   & Holmberg~II   &    08:19:04.0	& +70:43:09   & 7.9	& 3.39    &  7.8   & -16.7  &    9   &   0.116    & 0.7 & M~81        & 1.0  \\
UGC~4426   &  DDO~52        &    08:28:28.4	& +41:51:24   & 2.0	& 10.28  &  6.0   & -15.2  &  10   &   0.099   & -1.2 & NGC~2841  &  -0.9\\ 
UGC~4459   &  DDO~53        &    08:34:07.2	& +66:10:54   & 1.5	& 3.56    & 1.5    & -13.1  &  10   &   0.103    & 0.8 & M~81         & 1.0 \\
UGC~5139   &  Holmberg~I   &    09:40:32.3	& +71:10:56   & 3.6	 & 3.84   &  4.0   & -13.9  &  10   &   0.138    & 1.7 & M~81         & 1.8 \\
NGC~2976    &  UGC~5221  &  09:47:15.3	& +67:55:00   & 5.9	 & 3.56   &  6.1   & -16.7  &    7   &   0.195    & 2.9 & M~81         & 3.0  \\
UGC~5456    &                      &   10:07:19.6	& +10:21:46   &  1.6	  & 3.80  &  1.8   & -14.3  &    9   &   0.110    & -1.9 & NGC~2903  &   -1.3\\ 
UGC~5666    &    IC~2574    & 10:28:21.2	& +68:24:43   &	 13.2  & 4.02  & 15.4   & -17.3  &   8   &   0.099    & 1.0 & M~81         & 1.2  \\
UGC~6456    &                       &    11:27:59.9     & +78:59:24   & 1.6    & 4.3     &  2.0   & -13.7  &  10   &   0.102    & -0.1 & M~81        &  0.1  \\ 
UGC~6541    & Mrk~178        &    11:33:28.9	& +49:14:14   &	 1.2	  & 3.89  &  1.3   & -13.6  &   11   &   0.052    & -0.6 & M~81        & -0.2  \\
NGC~3738    &   UGC~6565  & 11:35:48.8	& +54:31:26   &  2.5	  & 4.90  &  3.6   & -16.5  &  10   &   0.028    & -0.8 & M81        & -0.4  \\
NGC~3741    &   UGC~6572  & 11:36:06.2	& +45:17:01   &	 2.0	  & 3.19  &  1.8   & -13.1  &  10    &   0.067   & -0.7 & M~81        & -0.4  \\
UGC~6817    &  DDO~99      &  11:50:53.0	& +38:52:49   &  4.1	  & 2.64  &  3.1   & -13.6  &  10    &   0.072   & -0.8 & NGC~4214  &  -0.4\\ 
NGC~4068    &  IC~757        &  12:04:00.8	& +52:35:18   & 3.3	 & 4.31   &  4.1   & -15.2  &    9   &   0.059    & -0.5 & NGC~4736  &  -0.1\\  
NGC~4163    &  UGC~7199  &  12:12:09.1	& +36:10:09   &	 1.8	  & 2.96  &  1.5   & -13.6  &    9    &   0.055   & 2.3 & NGC~4214   &  2.3  \\
UGCA~276    &  DDO~113   &  12:14:57.9	& +36:13:08   &  1.5	  & 3.18  &  1.4   & -11.8  &   10   &   0.054    & 1.3 & NGC~4214  &  1.5  \\
NGC~4214     &  UGC~7278 &  12:15:38.9	& +36:19:40   &	 8.5	  & 2.92  &  7.2   & -17.1  &    8    &   0.072    & 1.2 & NGC~4163  &   1.2 \\
UGC~7408     &   DDO~120  & 12:21:15.0     & +45:48:41   &  2.7   & 6.99   &  5.5   & -16.0  &    9    &   0.032   & 0.3  & NGC~4258   &  0.5  \\
NGC~4395     &  UGC~7524 & 12:25:48.9	& +33:32:48   &	 13.2  & 4.61   & 17.7  & -17.7  &    8    &   0.047   & 0.1 & NGC~4736  &   0.3 \\
UGC~7577     &   DDO~125    & 12:27:40.9	& +43:29:44   &	 4.3	  & 2.74   &  3.4   & -14.4  &   9    &   0.067   & -0.8 & NGC~4214  &  -0.4  \\
UGC~7605     &                       &  12:28:38.7	& +35:43:03   &	 1.1	  & 4.43   &  1.4   & -13.5  &  10   &   0.040   & 0.4 & NGC~4244   &  0.8 \\
NGC~4605     & UGC~7831     &  12:40:00.3	& +61:36:29   & 5.8	  & 5.47    &  9.2  & -17.8  &    8   &   0.039   & -1.1 & M~81       & -0.6 \\
UGC~7866     &   IC~3687      & 12:42:15.1	& +38:30:12   &	 3.4	  & 4.57   &  4.5   & -14.6  &  10   &   0.055   & 1.4 & NGC~4736    & 1.4  \\
UGC~8091    &  DDO~155    &  12:58:40.4	& +14:13:03   &	 1.1	 & 2.13   &  0.7   & -12.0  &  10   &   0.071    & -1.4 & MW         & -0.8 \\
UGC~8320     &   DDO~168    &  13:14:27.9	& +45:55:09   &	 3.6	  & 4.33   &  4.5   & -15.1  &  10   &   0.042   & 0.3 & NGC~4736  &   0.4  \\
NGC~5204     & UGC~8490    &  13:29:36.2	& +58:25:06   &	 5.0	  & 4.65   &  6.8   & -16.6  &    8   &   0.034   & -0.9 & NGC~4736  &  -0.4 \\
NGC~5238     & UGC~8565    &  13:34:42.5     & +51:36:49   &  2.0   & 4.21    &  2.4  & -14.3  &   10   &   0.027   & -0.4 & NGC~4736  &  -0.2 \\
UGC~8638     &                        &  13:39:19.4	& +24:46:32   &	 1.2	  & 4.27    &  1.5  & -13.1  &  10   &   0.036   & -0.2 & NGC~4826  &  -0.1 \\
UGC~8651     &  DDO~181      &  13:39:53.8	& +40:44:21   &	 2.3	  & 3.02    &  2.0  & -13.0  &  10   &   0.017   & -1.1 & NGC~4736  &  -0.6  \\
UGC~8760     &  DDO~183      &  13:50:50.6	& +38:01:09   &	 2.2	  & 3.24    &  2.1  & -13.1  &  10   &   0.045   & -1.0  & NGC~4736  &  -0.6  \\
NGC~5477     & UGC~9018     &  14:05:33.3     & +54:27:4     & 1.9     & 6.76    &  3.7  & -15.0  &    9   &   0.030   & 1.4  & M~101      &  1.4  \\
UGC~9240     & DDO~190        &  14:24:43.4	& +44:31:33   & 1.8	  & 2.80    &  1.5  & -13.9  &    9   &   0.034   &-1.2 & M~81       & -0.7  \\
NGC~6503     & UGC~11012    &   17:49:27.1	& +70:08:40   & 7.1	  & 5.27    & 10.9  & -17.8  &   6    &   0.088   & -1.1 & NGC~6946 &   -0.8\\  
NGC~6789     & UGC~11425    & 19:16:42.1     & +63:58:17   & 1.80   &  3.0     &  1.6  & -14.3  &    9    &  0.183    & -1.3 & M~81     &   -0.8  \\
UGC~12613   & Pegasus Dw   &   23:28:36.2	& +14:44:35   &  5.0	  & 0.97     &  1.1  & -12.1  &  10   &   0.187   &  0.9 & M~31     &    1.0  \\
 \hline         
\hline
  \end{tabular}
  \begin{tablenotes}
\small
      \item Columns~(1) and(2): galaxy name; 
      Col. (3) and (4): equatorial coordinates for the epoch J2000; 
      Col. (5): from K13, major angular diameter in arcminutes, corresponding to the 
      Holmberg isophote ($\sim$26.5 mag/arcsec$^2$) in the B band;   
      Col. (6): distance in Mpc; references to the distance estimates can be found in K13; 
      Col (7): linear major diameter in kpc, computed from the angular diameters in Col(5) and the distances in Col(6); 
      Col (8): absolute magnitude of the galaxy in the B band, computed adopting the apparent B magnitudes in K13 and the distances in Col. (6); 
      Col (9): morphological type according to the classification of \citet{devac91}; Col (10): Galactic extinction in the V band according to \citet{extinct}; 
      Col (10): tidal index $\theta_1$, as defined in K13 and Eq. (1) of this paper; Col (11): main disturber galaxy; Col (13): tidal index $\theta_5$, as defined in K13 and Eq. (2) of this paper.
     \end{tablenotes}
 \end{table*}

The targets for the SSH survey were selected from the Updated Nearby Galaxy Catalog  by \citet{k13}, hereafter K13, which consists of 
$\sim$900 nearby galaxies having individual distance estimates within $\sim$11 Mpc.  The K13 catalog collects several observables for each galaxy, including the angular diameter, the apparent magnitudes in different bands (far-UV, B, and K$_s$), the H$\alpha$ and H~I fluxes, the morphological type T as defined by \citet{devac91}, the radial velocity and the distance from one or more methods (e.g., TRGB, Cepheids, surface brightness fluctuations, Tully-Fisher and Fundamental Plane relations). The catalog also provides different indices that characterize the local environment,  such as the ``tidal'' indices $\theta_1$,  $\theta_5$, on which we will come back later. Starting from the K13 catalog, the SSH targets were selected according to the following criteria:
i) galaxies located in the northern hemisphere;  ii) with distance determination from the TRGB; iii) with foreground extinction $A_V<0.5$; iv) with morphological type T$\ge$6 (late spirals and irregulars); v) with absolute magnitudes in the range $-18<M_B<-11$;  vi) with apparent major diameter in the range 1-15 arcmin;  vii) with HST photometry available in the archive. These selection criteria resulted into a sample of 58 galaxies  (irregulars, blue compact dwarfs and  spiral dwarfs), from which we selected a smaller sub-sample of 45 galaxies uniformly distributed in right ascension to be observed with the LBT. Among the 13 galaxies not observed by SSH are objects already studied in the literature through deep, optical wide-field imaging, such as, e.g. NGC~4449  \citep{Martinez12}, Sextans B \citep{Bellazzini14}, KDG~61 and NGC~3077 \citep{Okamoto15}.

The adopted selection criteria are driven by the need to construct a sample large enough to allow for sufficient statistics in the search of rare, faint signatures of accretion events, and to permit a statistical characterization of the hierarchical formation process over a wide range of galaxy masses and  local environments. 
At the same time it is necessary to select targets for which the LBC, with a field of view of $\sim$23\arcmin$\times$23\arcmin, is most efficient in revealing the presence of nearby companions and substructures: we translated this into the requirement of a maximum angular diameter of $\sim$15\arcmin, to allow the entire galaxy and a significant portion of its surroundings to be imaged with one single LBC pointing, and a minimum angular diameter of $\sim$1\arcmin, below which even the small HST field of view can accomplish the goal. 
Ancillary HST data for all the galaxies included in the sample are needed to get deep and accurate photometry of the resolved stars 
that allow to infer high-quality SFHs; this requirement is driven by our goal to study the connection between accretion phenomena and starbursts in dwarf galaxies, as described in Section~\ref{section_objectives}. For the same reason, we restricted our sample to late-type systems, for which it is possible to reconstruct the recent SFH in great details. Our morphological selection naturally implies a bias toward low-density environments since, according to the well-known morphology-density relation \citep{Dressler80,Binggeli87,Binggeli90}, early-type dwarfs are typically located in the proximity of massive hosts while late-type dwarfs preferentially reside at the outskirts of galaxy groups; however, we found that the selected galaxies span a wide variety of environments, as quantified by the $\theta_1$ and $\theta_5$ tidal indices introduced by K13. 

 \begin{figure}
\includegraphics[width=\columnwidth]{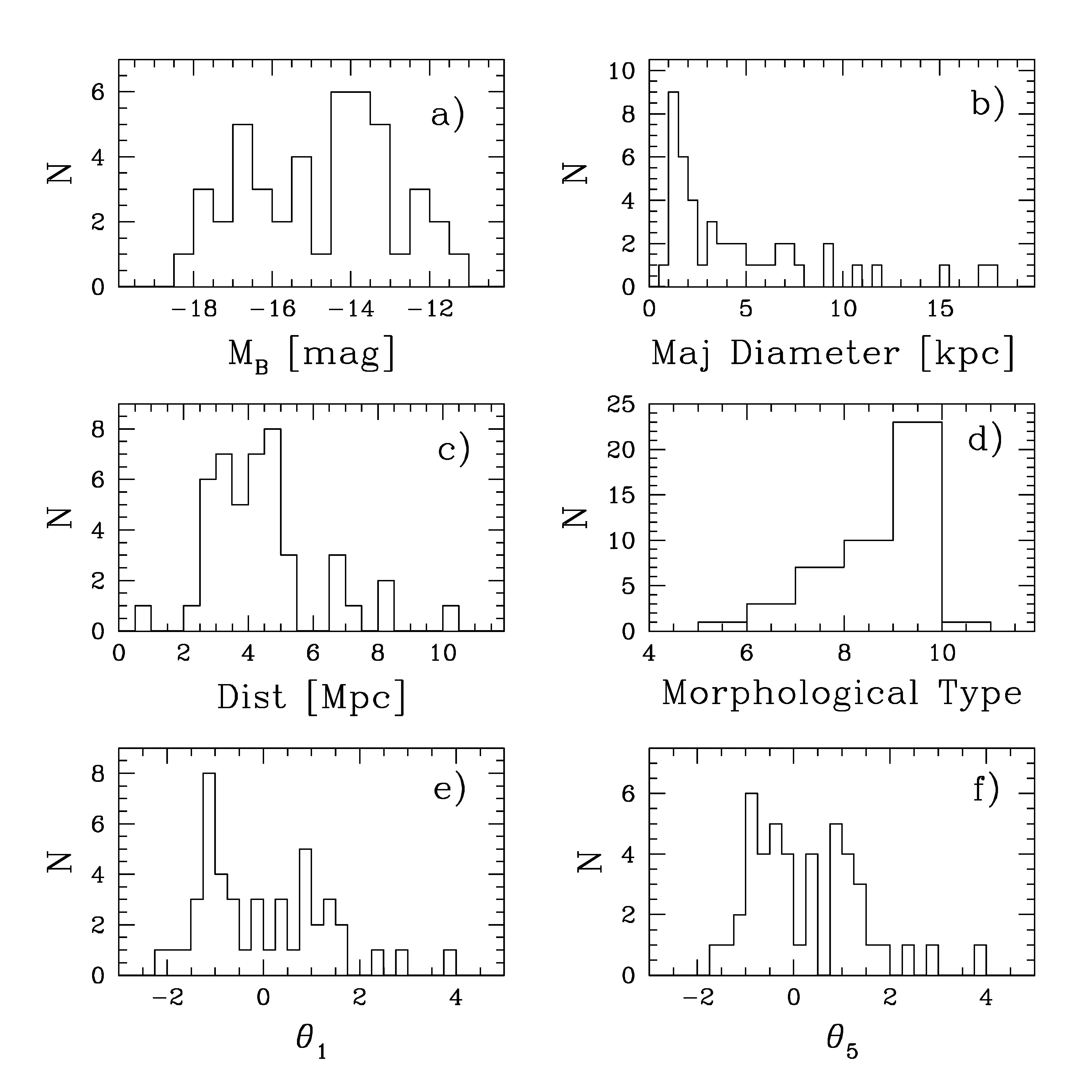}
    \caption{Properties of the 45 galaxies in the SSH sample. Distributions of a): galaxy absolute B magnitudes;
b): galaxy major diameters in kpc c): galaxy distances in Mpc; d): galaxy morphological types;  
 e) and f): tidal indices $\theta_1$ and  $\theta_5$, as defined by \citet{k13} (see Section~\ref{section_sample} for details); 
 galaxies with $\theta_1>$0 are members of a certain group, while galaxies with $\theta_1<$0 can be considered ``isolated''.}
    \label{sample_fig}
\end{figure}

More specifically, $\theta_1$ is defined as
\begin{equation}
\theta_1 = \log(M_{MD}/D_{MD}^3) + C
\end{equation}

\noindent where $M_{MD}$ and $D_{MD}$ are, respectively, the mass and the distance of the ``main disturber'' (MD) galaxy (i.e. the one inducing the highest tidal force $F\sim M/D^3$), and C is a constant chosen to give $\theta_1=0$ for a galaxy located on the ``zero velocity sphere'' relative to the MD. In practice, galaxies with  $\theta_1>0$ are members of a group, while galaxies with $\theta_1<0$ can be considered isolated. 

In order to account for the circumstance that $\theta_1$ can significantly change with time due to the orbital motion of galaxies, \cite{k13} also introduced the $\theta_5$ index, defined as the sum of the tidal forces produced by the five most important neighbors:
\begin{equation}
\theta_5 = \log \left(\sum_{n=1}^5 M_n/D_n^3  \right)  + C,
\end{equation}

\noindent which provides a more robust characterization of the tidal forces experienced by a galaxy over its lifetime. 
 
The final list of 45 galaxies in the SSH sample and their principal characteristics are presented in Table~\ref{sample_tab}. 
For each galaxy we provide the coordinates, the angular and linear major diameters, the distance derived from the TRGB, the absolute B magnitude, the morphological type T, the galactic extinction $A_V$, the two tidal indices $\theta_1$ and $\theta_5$, and the name of the main perturber galaxy.

\begin{figure*}
\includegraphics[width=\textwidth]{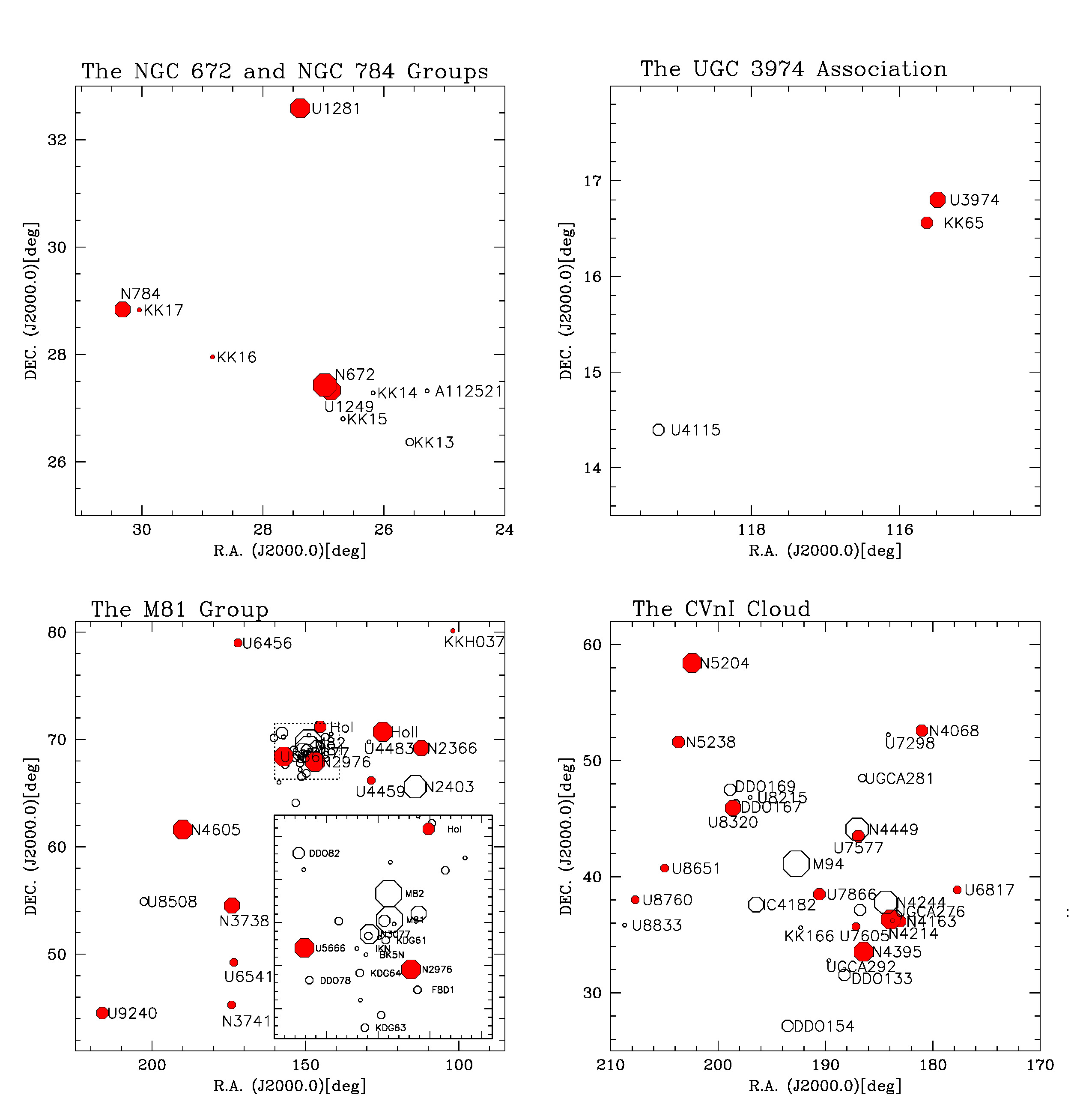}
    \caption{Spatial distribution of galaxies in the SSH sample belonging to known groups or associations: specifically, to the NGC~672/NGC~784 group \citep{k04,mcquinn14}, to the M81 group \citep[e.g.][]{chibo13}, to the UGC~3974 association \citep{tully06} and to the Canes Venatici I Cloud \citep{makarov13}. Symbol sizes are proportional to the galaxy brightness. Full red circles denote galaxies belonging to the SSH survey.}
    \label{groups_fig}
\end{figure*}

The properties of the sample in terms of absolute magnitude, linear size, distance, morphological type, and environment are presented in Figure~\ref{sample_fig}. All the magnitude bins in the range $-18<M_B<-11$ are well populated [panel a) of Fig.~\ref{sample_fig}], with the brightest galaxies reaching $\sim$twice the luminosity of the LMC. The galaxy major diameters are in the range $\sim$1-18 kpc, with a peak in the distribution at $\sim$1 kpc. 
The majority of the SSH galaxies are located between $\sim$2 and $\sim$6 Mpc, with only one target  as close as $\sim$1 Mpc and one as far as $\sim$10 Mpc (panel c). By construction, the morphological type  T is $\geq$6 (late spirals and irregulars), with a predominance of T$=$10 (Irr/Im) types. Finally, panels e) and f) show the distribution of the two tidal indices  $\theta_1$ and $\theta_5$ defined by \cite{k13} (see Equations (1) and (2)) describing the environment density. By definition, the indices increase with the ``stellar density contrast'' (either produced by the main disturber, as in the case of $\theta_1$, or due to the five most important neighbors, as in the case of  $\theta_5$). The galaxies in the SSH sample span tidal indices in the range 
$-2\lesssim \theta \lesssim2$, meaning that both galaxies at the center of groups and galaxies in isolation are present. At the two extremes we find UGC~1249 ($\theta_1=\theta_5=4.0$), which forms a group together with the spiral galaxy NGC~672 and other dwarfs, and the galaxy UGC~3755  ($\theta_1=-2.2$ and $\theta_5=-1.7$), 
which resides in the Lynx-Cancer Void. The majority of galaxies in the SSH sample belong to, or are in the vicinity of, known groups or associations; more specifically, there are five SSH targets (UGC~1249, UGC~1281, KK~16, KK~17, and NGC~784) in the NGC~672/NGC~784 group \citep{k04,mcquinn14}, thirteen (KKH~37, NGC~2366, UGC~4305, UGC~4459, UGC~5139, NGC~2976, UGC~5666, UGC~6456, UGC~6541, NGC~3738, NGC~3741, NGC~4605, UGC~9240) in the M81 group \citep[e.g.][]{chibo13}, two (UGC~3974, KK~65) in the UGC~3974 association \citep{tully06} and fourteen (NGC~4068, UGC~6817, NGC~4163, UGCA~276, NGC~4214, NGC~4395, UGC~7577, UGC~7605, UGC~7866, UGC~8320, NGC~5204, NGC~5238, UGC~8651, UGC~8760) in the Canes Venatici I Cloud \citep[CVn~I;][]{makarov13}. The spatial distribution of galaxies within these groups is shown in Fig.~\ref{groups_fig}. Galaxies with large positive values of $\theta_1$ or $\theta_5$ are found toward the center of the groups, while galaxies with highly negative $\theta$ values are peripheral. In addition, our sample comprises the Pegasus Dwarf (UGC~12613), which is a satellite of M~31; NGC~5477, associated to the M~101 spiral galaxy; and UGC~7408 in the CVn~II cloud.  The remaining eight galaxies (UGC~685, UGC~3755, UGC~4426, UGC~8091, UGC~5456, NGC~6503, NGC~8638, NGC~6789) are fairly isolated.

\section{Observations} \label{section_observations}

We were approved a two-year Strategic Program (PI F. Annibali) for the LBT Italian Cycle 2016$-$2017 to observe the 45 dwarf galaxies listed in Table~\ref{sample_tab}. Due to technical and environmental\footnote{Including the huge fire that burnt Mount Graham in June 2017} issues, only 25 of our targets were actually completed during the first two years;  we were then 
approved a carry-over of our program for the years 2018-2019. As of September 2019, data acquisition for  40 targets has been completed.

The SSH targets were observed with the wide-field ($23\arcmin\times23\arcmin$) Large Binocular Cameras, in binocular mode, in the  SDSS g and r passbands. The g and r images were obtained with the LBC-Blue and LBC-Red camera, respectively. 
The optics of each LBC feed a mosaic of four 4608~px $\times$ 2048~px CCDs, with a pixel scale of 0.225 arcsec~px$^{-1}$. Each CCD chip covers a field of $17.3\arcmin\times 7.7\arcmin$. Chips 1, 2, and 3 flank one another, being adjacent along their long sides; chip 4 is placed transversely to this array, with its long side adjacent to the short sides of the other chips \citep[see][]{giallo}. 

For each target, the total exposure time of 1 hr in each band was organized into 3 visits 
of 150 s $\times$ 8 dithered exposures or into 3 visits of 240 s $\times$ 5 dithered exposures, with the shorter exposure configuration preferred in presence 
of bright foreground stars contaminating the field. 
A variable pointing offset between the visits, depending on the angular size of the target galaxy, was implemented to make the galaxy light fall into different regions of the LBC chips, allowing for an optimal flat-field correction. 

We requested a seeing  $\lesssim$0.8\arcsec for targets at distances D$\lesssim$4 Mpc, and a seeing $\lesssim$1.2\arcsec for targets at larger distances. This difference reflects the distinct approaches (resolved star counts versus integrated light) selected for our search of faint companions, sub-structures, and stellar streams  around the SSH dwarfs depending on their distances. More specifically, a seeing better than $\sim$0.8\arcsec is needed to resolve individual stars down to $\sim$1 mag below the TRGB in galaxies within D$\sim$4$-$5 Mpc (depending also on the crowding). At larger distances, on the other hand, resolving individual RGB stars is no longer feasible from the ground in seeing limited mode,  even in the least-crowded galaxy outskirts; therefore, we relaxed our constraint on the seeing, but required dark moon observation, in order to keep the background low and be able to detect integrated-light overdensities in the g and r images.

For each galaxy, the 1 hr total exposure time per band is dictated by the need to measure stars down to $\sim$1 mag below the TRGB  at D$=$4 Mpc with a signal-to-noise $\gsim$5 in both $g$ and $r$.  At 4 Mpc, the TRGB tip is expected at r$\simeq$25.3,  g$-$r$\simeq$1 (for metal-poor populations), therefore we need to reach stars with $r\simeq26.3$, g$\simeq27.3$. The same exposure time was adopted also for targets closer than 4 Mpc in order to reach the same uniform apparent depth for all the sample. The impact of crowding is mitigated by the fact that we are mainly interested in resolving stars in the very low surface brightness external galaxy regions. 
The same 1 hr exposure time was also adopted for the  un-resolved part of the sample at D$>$4 Mpc, since previous studies with the LBT \citep[e.g.][]{Annibali16} 
have shown that this is sufficient to reveal integrated-light sub-structures as faint as  $\mu_r\sim29$ mag/arcsec$^2$ (or better, see Sect.~\ref{section_udg}). 

The choice of the $g$ and $r$ bands provides the best compromise to: a) be able to reach old ($>$1-2 Gyr) RGB stars, b) minimize, as far as possible, the contribution from the sky background, which becomes particularly important at wavelengths longer than the r band, and  c) calibrate the LBT photometry through g and r catalogs available from the Sloan Digital Sky Survey and/or PS1. Short 60 sec $\times$ 4 exposures in $g$ and $r$ were also obtained to get un-saturated photometry of bright foreground stars to be used to anchor the photometry to the SDSS system.

\begin{figure*}
\includegraphics[width=\textwidth]{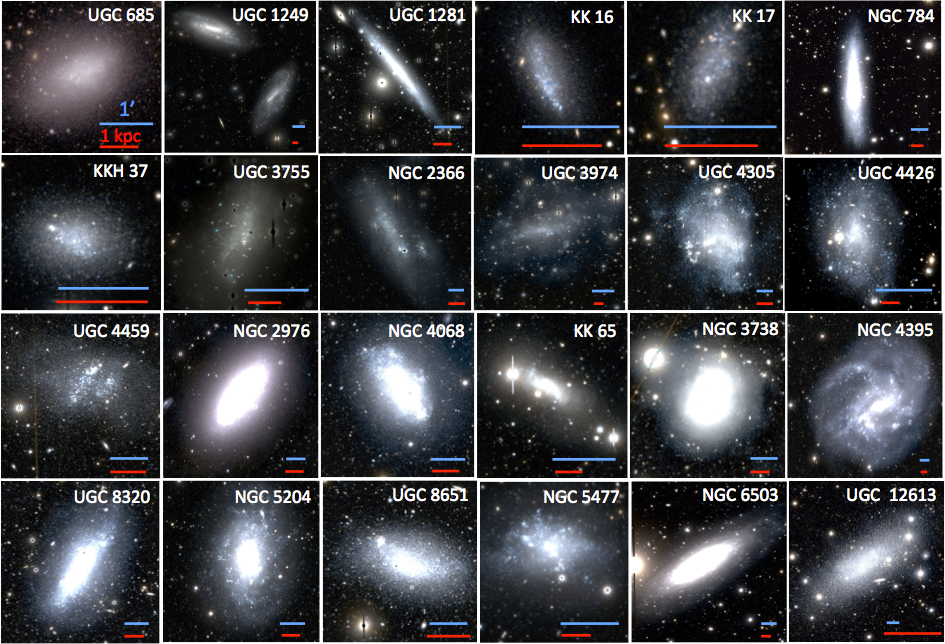}
    \caption{Montage showing portions of the LBC two-color ($g$ and $r$) images for the 24 SSH galaxies already analysed. In each panel we show the name of the galaxy and two bars showing the size of  1$\arcsec$ (upper bar, blue in the electronic version of this paper) and 1 kpc (lower bar, red). The physical size of each galaxy was computed adopting the distances  listed in Table~\ref{sample_tab}.}
    \label{montage}
\end{figure*}

\section{Image Reduction} \label{section_reduction}

Image reduction and creation of the final $g$ and $r$ stacked mosaics were performed  using a specific pipeline developed at INAF-OAR. The main steps of the procedure are described below. 

\subsection{Pre-reduction} 

First, the raw LBC images were pre-reduced removing a master-bias and applying a master-flat to normalize the response of each image pixel. 
The master-bias was obtained from the median stack of a set of bias frames. The master-flat is the stack of a set of twilight sky-flats acquired with the same observational setup of the science images, to which the bias was previously subtracted; furthermore, each individual flat was normalized to its own median background level, so that the final stacked 
master-flat is normalized to unity.  During pre-reduction, several image artifacts, such as saturated and hot/cold pixels, cosmic rays (quite frequent), and trails from orbiting 
satellites (rare) were also masked out. 
 
\subsection{Background Subtraction} 

Typically, after pre-reduction, the images are still far from being ``flat'', and both small- and large-scale structures, such as pupil ghosts, dust, and sky-background variation during the observations, may still be present in the images. To cope with this, background maps were subtracted to the pre-reduced images. The SExtractor package \citep{sextractor}
allows to construct background maps using a combination of k-$\sigma$-clipping and mode estimation to evaluate the local background 
in each mesh of a grid that covers the whole frame, and then performing an interpolation between the grid meshes. The resulting map crucially depends on the user-defined 
BACK\_SIZE parameter, which defines the size of the mesh. In particular, if the mesh size is too small, roughly comparable to the source spatial extension, a background over-subtraction at the source edges may occur. The larger the source, the more dramatic the over-subtraction.  On the other hand, if the mesh size is too large, the map will fail to reproduce the small-scale variations of the background. 
To deal with this difficulty, the following series of steps were implemented into the INAF-OAR pipeline: 1) first, SExtractor  was run to detect sources and to create a mask where the detections are flagged-out; 2) the mask was then transformed into a ``broadened'' version, where the masked regions were ``expanded''  to flag-out residual flux potentially present around the sources; 3) bad pixels and artifacts were also masked out; 4) SExtractor was run again with a relatively small mesh size and with the flagged-pixel mask to create the background map. The small mesh size allows us to follow the small-scale background variation, while the relatively extended masked regions at the source positions prevent to over-subtract the background at the source edges.

\subsection{Astrometric Solution}

The individual images present geometric distortions due to various effects  (optics distortions, atmospheric refraction, chip rotations); therefore, astrometric calibration is needed 
before co-adding the individual frames into a final stacked mosaic image. The astrometric calibration was accomplished through two separate steps: first, relative linear offsets between the individual exposures were derived through cross-matching of the source catalogs created by SExtractor with respect to a chosen reference frame; then, the absolute astrometric calibration was accomplished  through cross-matching onto a reference absolute astrometric catalog, such as the United States Naval Observatory (USNO). The astrometric solution was derived and stored into the image header. Note that in the final stellar catalogues this solution will be re-adjusted using a dense grid of stars in common with PanSTARRS1 (PS1), see Sect.~\ref{phot_astrometry}, below.

\subsection{Noise and Image Co-addition}

Noise maps were created from the individual images under the assumption that the noise follows the Poisson statistics of the counts measured in each pixel of the original frames. 
To evaluate the noise, the scaling factors applied to each pixel during processing (e.g. flat-fielding, exposure time normalization, etc.) were taken into account. More specifically, each noise map was computed according to the formula: 

\begin{equation}
\sigma(x,y)_i = \sqrt{\frac{counts_{RAW}(x,y)_i} {gain_i} \times \frac{1}{flat(x,y)_i}}
\end{equation}

\noindent  where $i=1,2,3,4$ denotes the chip number,  $\sigma(x,y)$ is the computed noise at the (x,y) position, $counts_{RAW}$ is the number of counts in the raw image, $gain$ is the read-out-gain ($\sim$2), and $flat$ is the value of the flat image used in the pre-reduction.  

The individual processed images were resampled according to the geometry described in a global header and finally they were combined into $g$- and $r$-band stacked images
making use of the  SWarp software \citep{swarp}. The noise maps  $\sigma(x,y)$ were used to derive the weights [$w_i(x,y)=1/\sigma(x,y)^2$] that enter in the weighted summation of all individual frames.

 As of today, image reduction has been completed for 24 targets in the SSH sample. Portions of their LBC $g$, $r$ color combined images are shown in Fig.~\ref{montage}.

\section{Photometry} \label{section_photometry}

Photometry of individual sources was performed on stacked mosaic images using PSFEx \citep{PSFEX}. In practice PSFEx produces a numeric model of the Point Spread Function (PSF) using the brightest isolated stars in each image. In the present case, the spatial variation of the PSF was modeled with a second order polynomial in the x and y image coordinates. Then the derived model is used by SExtractor to estimate PSF magnitudes by fitting the proper model PSF to each detected source, according to its position in the image.

The detection was made independently in both passbands, using rms images as weight maps. We used a mexhat filter and we considered as valid detections all the sources having a flux level larger than $3\sigma$ above the background in at least 3 
adjacent pixels. Even if the sky was subtracted from our images, we required it to be estimated locally, over 24~px thick annuli around each source, to remove any possible local residual of the sky subtraction. The catalogues from the g and r images where matched and combined using the Catapack suite\footnote{\tt http://davide2.bo.astro.it/~paolo/Main/CataPack.html}, as done in \cite{Bellazzini14,Bellazzini15}. In our catalog we include the following parameters provided by SExtractor and
PSFEx: an identification number, the position of the sources derived with the PSF fitting, both in image and world coordinates, aperture magnitudes over a diameter of $2.5\arcsec$ and the associated uncertainties, 
MAG\_BEST\footnote{see \citet{sextractor} and {\tiny \tt http://astroa.physics.metu.edu.tr/MANUALS/sextractor/Guide2source\_extractor.pdf} for a detailed description of the meaning of this as well as other output parameters of Sextractor.}, 
as a reasonable magnitude estimate for extended sources, the PSF magnitudes and the associated uncertainties and $\chi^2$ value, the SExtractor quality flag and stellarity index, the axis ratio, the half-light radius and the maximum surface brightness in mag/arcsec$^2$.

We remark that, while our catalog contains useful information on extended sources, the overall data reduction strategy has been optimized to get the best results and the most reliable photometry for point sources (stars).

\subsection{Astrometric and photometric calibration.} \label{phot_astrometry}

The astrometric and photometric calibration procedure follows strictly \citet{Bellazzini15}, where a detailed  description is provided. The only difference is that here we use catalogs extracted from PS1, instead of SDSS, to derive the photometric zero points and the final astrometric solution. We chose PS1 for homogeneity as all our targets are included in the PS1 footprint, while this is not the case for the SDSS. PS1 photometry is transformed into the SDSS photometric system using the equations provided by \citet{tonry}.

The original astrometry embedded in the images and transported into the original catalogs by SExtractor was transformed into the PS1 system with third order polynomials, using several hundreds of stars in common between our catalogs and the corresponding PS1 catalogs. In all cases, we carefully checked that the stars used to compute the transformation were distributed all over the LBC field. The typical rms of the solutions is $\le 0.1\arcsec$ in both directions.

The brightest well-measured stars in common were used to fit the photometric zero points, once color terms were removed from our instrumental magnitudes with Eq.~A.1 of \citet{Bellazzini15}.
The range of magnitude in which our deep photometry and PS1 photometry overlaps is limited on one side by saturation and on the other side by large photometric errors near the PS1 limiting magnitude, respectively.
For this reason, the uncertainty on the zero points may amount to $\sim 0.02-0.03$ magnitudes \citep[see][]{Bellazzini15}, the zero point in the r band being the most uncertain of the two. However, in all the cases, the CMD from the original PS1 matches very well the CMD from our calibrated photometry.

In the following we will use final catalogues in which are only sources having valid PSF magnitudes in both passbands, and we will refer to PSF magnitudes, if not otherwise stated.

  \begin{figure}
\includegraphics[width=\columnwidth]{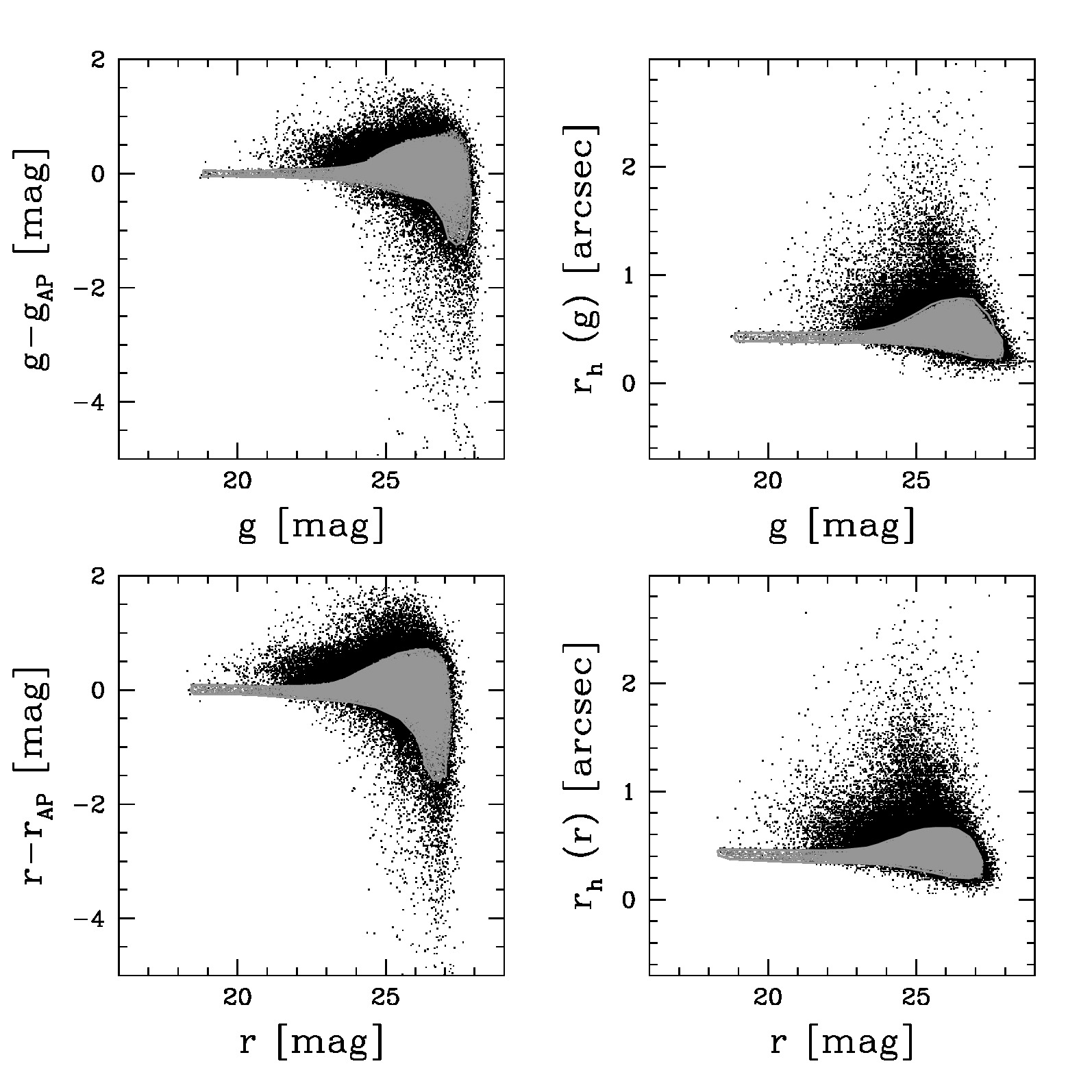}
    \caption{Our adopted selections on the photometric catalogs. {\bf Left panels:} PSF-fitting minus aperture magnitudes versus PSF magnitudes in g (top) and r (bottom). {\bf Right panels:} Half light radius versus magnitude in both g and r. Black points correspond to all stars with SExtractor flags equal to zero  (i.e. well measured sources, not blended to other objects or affected by the presence of bright neighbours) while grey points denote the adopted selections as described in Sect.~\ref{section_sel_cuts}. 
 }
    \label{phot_sel}
\end{figure}

\begin{figure*}
\includegraphics[width=\textwidth]{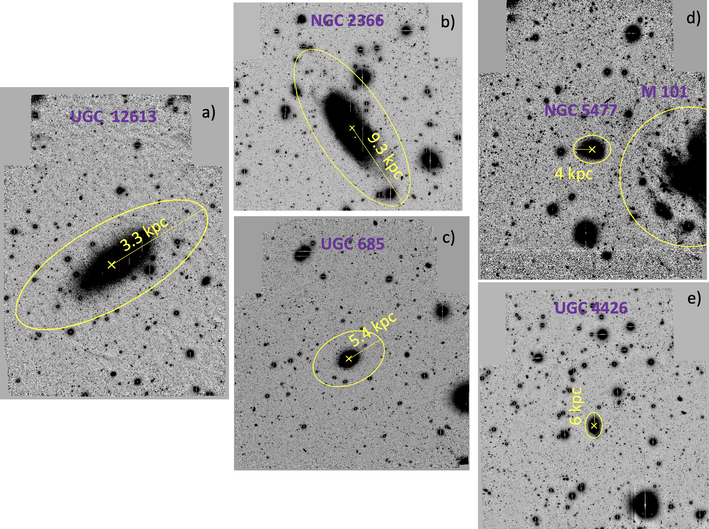}
    \caption{LBT LBC $g$ images of the five galaxies discussed in this paper: panel a), UGC~12613 (Pegasus dwarf irregular), at D=0.97 Mpc; panel b), NGC~2366, at D=3.2 Mpc; panel c), UGC~685, at D=4,7 Mpc; panel d), NGC~5477, at D=6.8 Mpc; panel e), UGC~4426, at D=10.3 Mpc. The images show the entire observed field of views, with the display cuts chosen to emphasize the outer low surface brightness regions. For each galaxy, the ellipse indicates the region within which  we are able to identify, from the $g$, $r$ CMD, stars belonging to the galaxy; outside the ellipse, the CMD  appears populated by only background galaxies and foreground MW stars. The ellipse semi-major axis in kpc is also indicated.}
    \label{five_gals}
\end{figure*}

\subsection{Selection Cuts} \label{section_sel_cuts}

In order to remove spurious detections from the final photometric catalogs, we applied selection cuts using some of the SExtractor output quantities. 
In the first place, we performed a selection based on the  flag parameter, which provides a record of all possible problems encountered during the photometry 
\citep[see][]{Holwerda05}.  In particular, flag=1 indicates the presence of bright nearby neighbours that may have affected the source photometry, flag=2 indicates that the source was originally blended to another object, and higher  flag values denote the presence of saturated pixels, truncated objects, and corrupted or incomplete output data or combination of these problems. 
Retaining all sources with flag=0 provides the most conservative and cleanest selection of well-measured, individual sources, and we adopted this selection for the closest galaxies in our sample, at D$<$5 Mpc, where we expect to resolve  (depending also on the seeing and on the stellar crowding) individual stars down to the RGB level. On the other hand, we relaxed our constraints for more distant galaxies and selected all sources with flag=0,1,2 in order to at least retain bright young stars, located in regions of active star formation and typically blended to other sources. This is the only kind of selection that was applied to targets at D$\ge$5 Mpc in our sample.  

For targets at  D$<$5 Mpc, further selection cuts on parameters related to the source's size were applied to remove extended background galaxies from the catalogs; 
an exploration of all possible combinations of the SExtractor output parameters indicates 
that selections on the distributions of the PSF-fitting minus aperture magnitudes ($g-g_{ap}$, $r-r_{ap}$)  or of the half-light radius ($r_h(g,r)$) provide the most effective removal of extended sources. As an illustrative example, we show in Fig.~\ref{phot_sel} the selection adopted for UGC~12613 (Pegasus dwarf galaxy). The left panels exhibit the distribution  of $g-g_{ap}$ versus $g$ and  $r-r_{ap}$ versus $r$  for all objects with flag($g,r$)=0; the distribution resembles that of the sharpness versus magnitude typically obtained from other widely-used PSF-fitting 
photometric packages \citep[e.g. Daophot,][]{Stetson87}. At bright magnitudes, the distribution peaks around zero, as expected for point-like sources. An   additional broad peak at $g-g_{ap}$, $r-r_{ap}$$\sim$0.2 is visible as well; the presence of this feature is due  extended objects (background galaxies, stellar clusters, blends of two or more stars), for which the PSF does not provide a good description of the profile and thus underestimates the source flux. 
 At the opposite extreme, toward faint magnitudes, low $g-g_{ap}$, $r-r_{ap}$ values indicate sizes smaller than the PSF, likely associated to cosmic rays or detector's artifacts. Following these considerations, we adopted the selections provided by the grey shaded area in the left panels of Fig.~\ref{phot_sel}. The other SExtractor parameter that turns out useful to separate stars from background galaxies is the half-light radius $r_h(g,r)$, whose distribution as a function of magnitude in both bands is shown in the right panels of Fig.~\ref{phot_sel}. Extended objects removed on the basis of the PSF minus aperture magnitude distributions exhibit also large values of $r_h(g,r)$. A selection in the  $r_h(g)$ vs. $g$ and $r_h(r)$ vs. $r$ planes, as shown in the right panels of  Fig.~\ref{phot_sel}, allows for a refined removal of the extended sources that were not captured by the previous selections.

\begin{figure*}
\includegraphics[width=\textwidth]{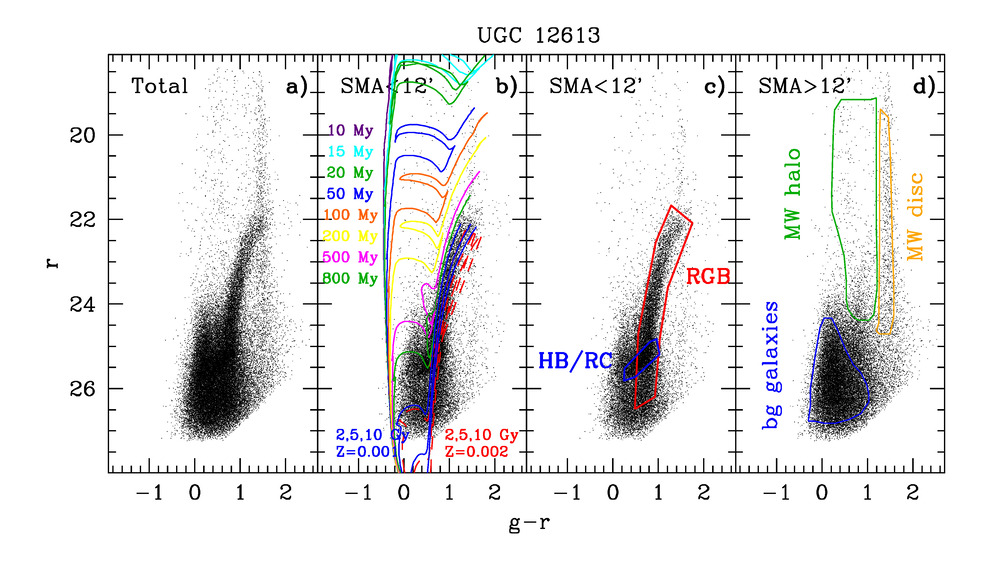}
    \caption{$r$, $g-r$ color-magnitude diagram (CMD) of UGC~12613 (Pegasus dwarf). a): total CMD over the whole 
    $23\arcmin\times23\arcmin$ LBC field of view. 
    b): CMD of stars within an ellipse of ellipticity $\epsilon=0.6$ and semi-major axis (SMA)=12$\arcmin$; overplotted are the PARSEC \citep{Bressan12} isochrones in the SDSS photometric 
  system for different ages,  shifted to a distance of D=0.97 Mpc [(m-M)$_0$=24.9] and corrected for a foreground reddening of E(B-V)=0.06 \citep{extinct}. The displayed Z=0.002 isochrones roughly correspond to the metallicity measured in UGC~12613' s H~II regions \citep[12+log(O/H)=7.9$\pm$0.1-0.2, i.e. $\sim$1/7 solar,][]{skill97}; for ages $\ge$2 Gyr, we also show the 
    Z=0.001 ($\sim$1/16 solar) metallicity models. For clarity, the thermally-pulsing asymptotic giant branch (AGB) models have not been displayed in the plot. c): same as b), but with 
     indicated the RGB and the RC/HB phase. d): CMD of stars at SMA$>12\arcmin$, mostly sampling background galaxies, and foreground MW halo and disc stars, as labelled in the plot.}
    \label{cmd_u12613}
\end{figure*}

\section{Color-Magnitude Diagrams} \label{section_cmd}

To analyse which kinds of populations have been resolved by our photometry, we plotted the CMDs of all the SSH fields and compared them with theoretical isochrones. In several cases, the CMD of the total field of view is heavily contaminated by foreground Milky Way stars (specially if the target is at low Galactic latitudes), either disc or halo ones, and by background galaxies. For this reason, the CMDs of more internal galactic regions are often more informative on the evolutionary properties of the target galaxy than the global ones. We have also checked if the CMD evolutionary sequences show trends with galacto-centric distance (i.e., gradients) or with other position parameters (i.e., signs of perturbations). 

We consider it useful to illustrate the kind of CMDs that we obtain with SSH, and the corresponding analyses,  in this survey presentation paper. We thus show the CMDs for a few representative cases in the SSH survey: UGC~12613, NGC~2366, UGC~685, NGC~5477, and UGC~4426, located at distances of $\sim$1, 3.2, 4.7, 6.8, and 10.3 Mpc, respectively (see Fig.~\ref{five_gals}). In particular, UGC~12613, and  UGC~4426 are the closest and the most distant galaxy in our sample, respectively.  
These five illustrative examples provide a direct grasp of our ability in resolving individual stars as a function of galaxy distance. 
Previous studies have demonstrated that individual RGB stars can be resolved from the ground in 
the least crowded galaxy outskirts up to distances of $\sim$4-5 Mpc, provided that the seeing is as good as $\lesssim$ 0.8\arcsec \citep[e.g][]{Bernard12,Okamoto15,pisces}.  
The CMDs obtained from our analysis confirm these results, showing that RGB stars, tracing the underlying old ($>$1-2 Gyr) stellar population, are reached up to 
targets at D$\lesssim$5 Mpc, while only brighter stellar phases (young main sequence stars, blue and red supergiants, asymptotic giant branch stars)  are accessible at larger distances.

A detailed description of the CMDs for all five targets follows below.    

 \begin{figure*}
\includegraphics[width=\textwidth]{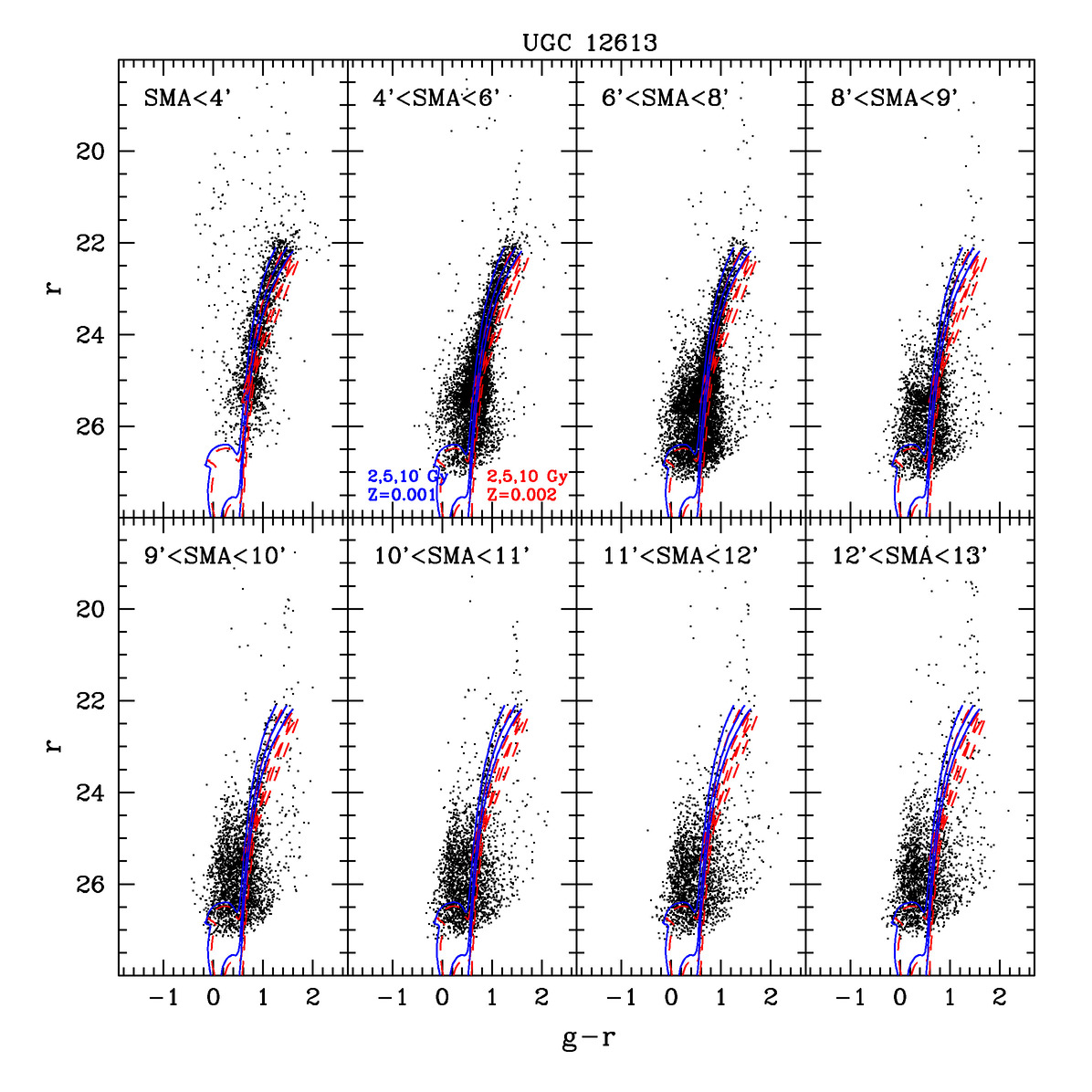}
    \caption{CMDs of UGC~12613 selected in elliptical annuli of increasing semi-major axes (SMA), where 1 arcmin corresponds to $\sim$0.3 kpc at the galaxy distance. Overplotted are the RGB portions of the PARSEC \citep{Bressan12} isochrones in the SDSS photometric system shifted to a distance of D=0.97 Mpc ((m-M)$_0$=24.9)  and corrected for a foreground reddening of E(B-V)=0.06. The dashed (red in the electronic version of the paper) isochrones have metallicity  Z=0.002, whilst the solid ones (blue) have Z=0.001. For clarity, the thermally-pulsing AGB models have not been displayed in the plot.
   \label{u12613_cmd_radial}}
\end{figure*}

\begin{figure*}
\includegraphics[width=\textwidth]{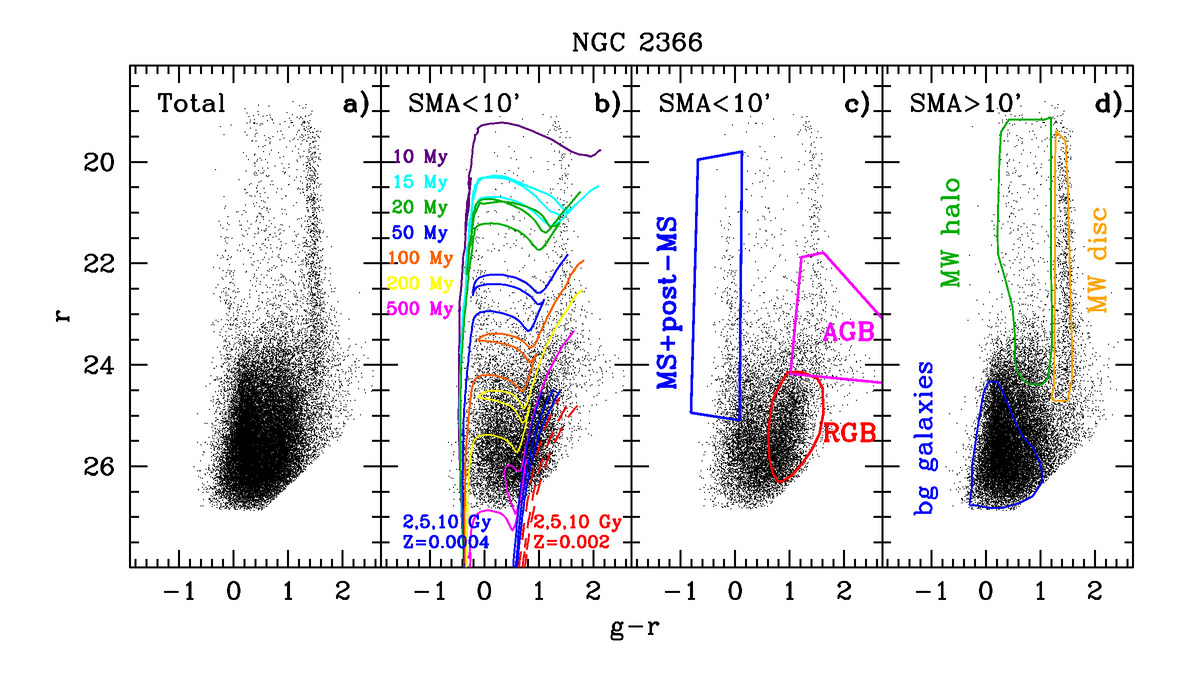}
    \caption{$r$, $g-r$ CMD of NGC~2366. a): total CMD over the whole 23\arcmin$\times$23\arcmin  LBC field of view. 
    b): CMD of stars within an elliptical region of semi-major axis (SMA) = 10\arcmin; overplotted are the PARSEC \citep{Bressan12} isochrones in the SDSS photometric 
    system for different ages, shifted to a distance of D=3.2 Mpc [(m-M)$_0$=27.5] and corrected for a foreground reddening of E(B-V)=0.03 \citep{extinct}. The displayed isochrones have a metallicity of Z=0.002, roughly consistent with the NGC~2366 H~II regions' metallicity \citep{Izotov97,Noeske00,James16}; for ages $\ge$2 Gyr, we also show the 
   Z=0.0004 ($\sim$1/40 solar) metallicity models. For clarity, the thermally-pulsing AGB models have not been displayed in the plot. c): same as b), but with 
     indicated some relevant stellar evolutionary features: the solid blue line encircles the bulk of MS and post-MS stars  at the hot edge of their core helium  burning phase;  the magenta line encircles the bulk of asymptotic giant branch stars (AGB), while the red line the bulk of RGB stars. c): CMD of stars at SMA$>10\arcmin$, mostly sampling background galaxies, and foreground MW halo and disc stars, as labelled in the plot. }
    \label{cmd_n2366}
\end{figure*}

\subsection{UGC~12613}

UGC~12613 (also known as the Pegasus dwarf irregular galaxy), at a distance of  D=0.97 Mpc \citep{EDD}, is the closest target in our sample. 
Its final CMD, obtained after application of the selection cuts described in Section~\ref {section_sel_cuts}, is shown in Fig.~\ref{cmd_u12613}. There we display both the total CMD (panel a) obtained from all sources detected in the  
$23\arcmin\times23\arcmin$ LBC field of view, and two selections respectively for sources within an elliptical contour with ellipticity 
$\epsilon=0.55$, major axis position angle PA$=125^{\circ}$ (see Section~\ref{section_profiles} below) 
and semi-major axis (SMA) = $12\arcmin$ (panels b and c)\footnote{Here we denote as SMA the elliptical radius as defined in \citet{Bellazzini11}.}, as indicated in Fig.~\ref{five_gals}, and for sources outside that region (panel d). 

The external field at SMA$>12\arcmin$ provides  a representation of the foreground and background contamination affecting our CMD of UGC~12613:  as discussed in details by \cite{Bellazzini14}, the narrow vertical feature at $g-r\sim1.5$ is due to the population of local M dwarf stars, while the sparse band of stars extending from $g-r\sim1$,  $r\sim24$ to $g-r\sim 0.2$, $r\sim19$ is made of main sequence stars at different distances in the MW halo. On the other hand, the triangle-shaped clump of objects at $0\leq g-r \leq1$,  $r\ge 24$ is due to unresolved background galaxies. The presence of these contaminants must be taken into account when discussing the CMD of the internal field.

The CMD of UGC~12613 at SMA$<12\arcmin$ exhibits a well defined and narrow ($\Delta(g-r)\sim0.3$ mag) red giant branch,
 extending from $g-r\sim1.5$, $r\sim22$ down to $g-r\sim0.6$, $r\sim25$, populated by stars older than 1 Gyr, and 
the fainter sequence of the red clump (RC) or horizontal branch (HB) at $0\lesssim g-r \lesssim1$,  $25\lesssim r \lesssim26$,  corresponding to low-mass core helium burning stars, from a few Gyr to 13 Gyr old. 
 Younger stars are  scarse in the CMD, likely because confined to the most internal galaxy regions where severe crowding hampers their detection.  In panel b) of Fig.~\ref{cmd_u12613}, the PARSEC stellar isochrones \citep{Bressan12} have been  superimposed  to the observed CMD adopting a distance of 
 D=0.97 Mpc  ((m-M)$_0$=24.93) from \cite{EDD} and correcting for a foreground reddening of E(B-V)=0.06 \citep{extinct}. Internal reddening can be neglected here, given that dust extinction is found to be modest in low metallicity galaxies and typically confined to regions of active star formation \citep[see e.g.][]{draine07,madden13}; indeed our photometry is strongly incomplete toward the most central, star-forming regions, and the vast majority of detected RGB or RC/HB stars reside in external, close to dust-free, galaxy zones.  
 The displayed models cover a wide age range from 10 Myr to 10 Gyr, and have a metallicity of Z=0.002, consistent with the metallicity measured in UGC~12613' s H~II regions \citep[12+log(O/H)=7.9$\pm$0.1-0.2, i.e. $\sim$1/7 solar,][]{skill97}. However, it is immediately clear that these models predict too red RGBs, indicating that the stars older than 1$-$2 Gyr are more metal poor than the present-day interstellar medium.   
 Indeed, models as metal poor as Z=0.001 ($\sim$1/16 solar metallicity), displayed in Fig.~\ref{cmd_u12613}  provide a more satisfactory fit to the RGB colors.  
 
 There has been some debate in the literature about the distance of the Pegasus dwarf irregular \citep[e.g.,][]{Aparicio94,Galla98}, with the most recent estimates providing values in  the range 0.9$-$1.2 Mpc \citep{Mc05,Piet10,McCall12,Tully13,McQuinn17}. Our dataset provides the most reliable detection of the RGB tip so far,   thanks to  the large number of measured stars, from which we derive a distance modulus of $(m-M)_0=24.94$, or D=0.97 Mpc, compatible with the quoted range.

 The behaviour of the stellar populations in UGC~12613 as a function of galacto-centric distance is illustrated in Fig.~\ref{u12613_cmd_radial}, where we plot the CMDs for stars selected 
 in elliptical annuli of increasing semi-major axes (SMA). Overplotted to the observed CMDs are the PARSEC stellar isochrones for ages$\ge$2 Gyr, and for the Z=0.001 and Z=0.002 metallicities. Stars younger than $\sim$1 Gyr are only detected within the internal SMA$<4\arcmin$ region ($<$1.2 kpc), although we caution that some of these sources may in fact be unresolved blends of two or more individual stars or compact star clusters. The RGB and HB/RC populations due to the galaxy can be recognized in the CMDs out to SMA$\sim12\arcmin$ ($\sim$3 kpc), while at larger radii it is only possible to recognize in the CMD the contribution due to background galaxies and MW stars. 
No significant metallicity variation with increasing galacto-centric distance is observed in the CMDs, and the RGB colors are overall compatible with a Z=0.001 metallicity.

\begin{figure*}
\includegraphics[width=\textwidth]{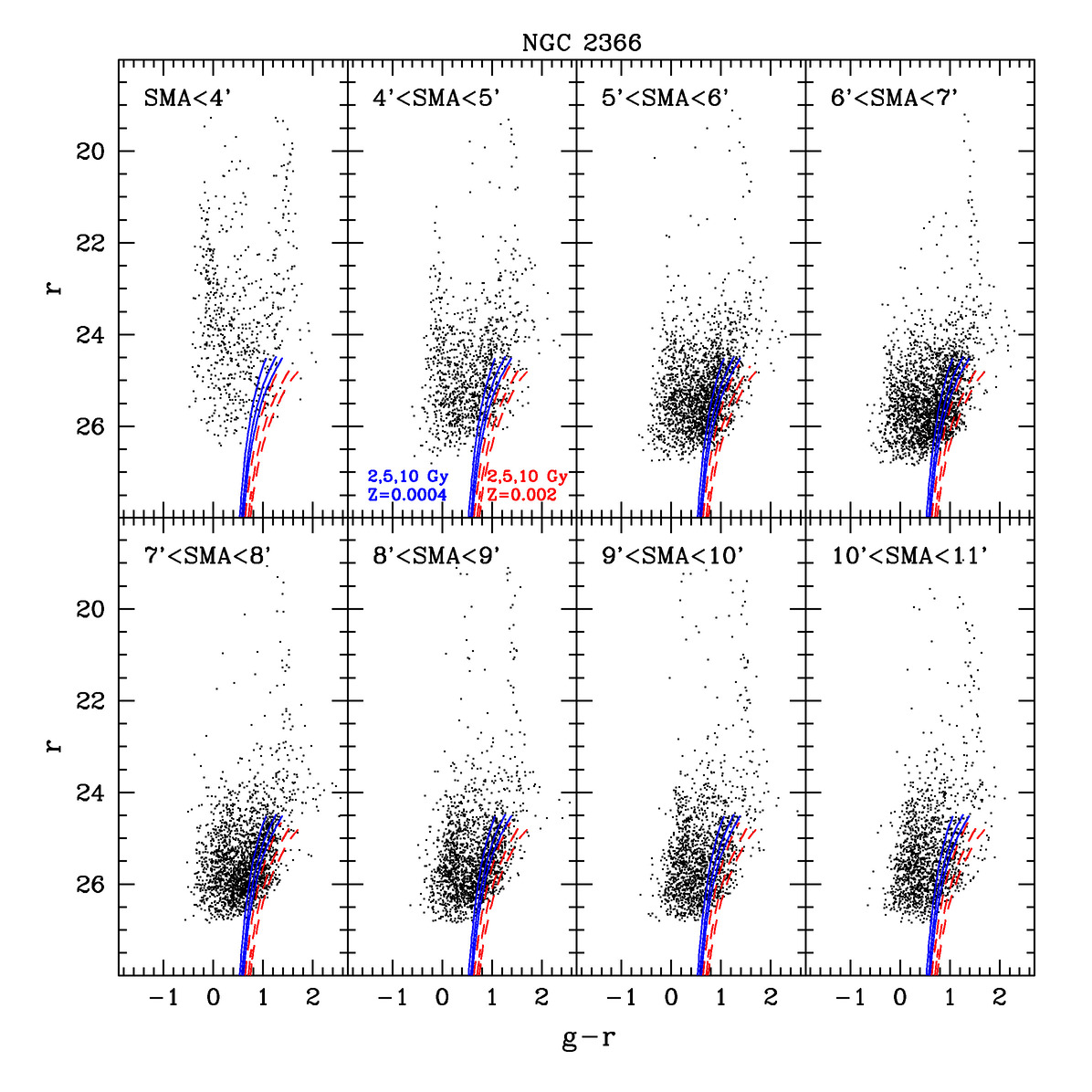}
    \caption{CMDs of NGC~2366 selected in elliptical annuli of increasing semi-major axes (SMA). Overplotted are the RGB portions of the PARSEC \citep{Bressan12} isochrones in the SDSS photometric system shifted to a distance of D=3.2 Mpc ((m-M)$_0$=27.5). The dotted (red in the electronic version of the paper) isochrones have metallicity  Z=0.002, whilst the solid ones (blue) have Z=0.0004. For clarity, the thermally-pulsing AGB models have not been displayed in the plot. 
    }
    \label{n2366_cmd_radial}
\end{figure*}

\begin{figure*}
\includegraphics[width=\textwidth]{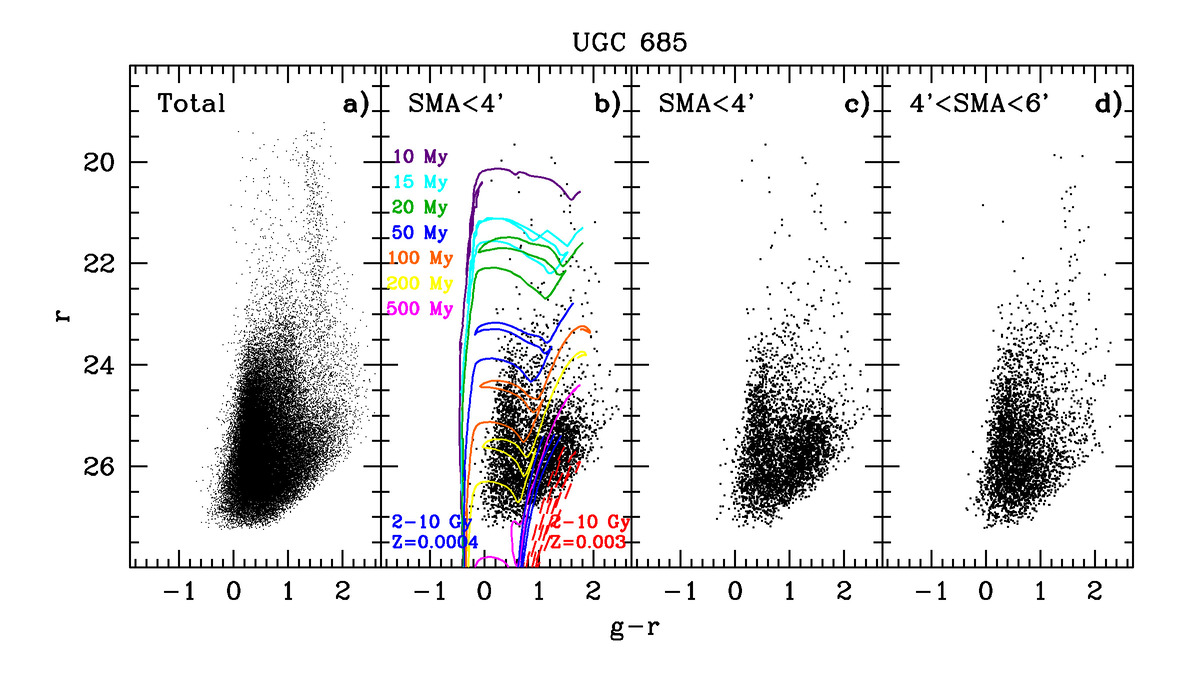}
    \caption{$r$, $g-r$ CMD of UGC~685. a): total CMD over the whole 23\arcmin$\times$23\arcmin LBC field of view. 
    b): CMD of stars within an elliptical region of semi-major axis (SMA) = 4\arcmin, with overplotted the PARSEC isochrones shifted to a distance of D=4.7 Mpc [(m-M)$_0$=28.4] and corrected for a foreground reddening of E(B-V)=0.05. The displayed isochrones have a metallicity of Z=0.003, compatible with the H~II regions metallicity of 12 + log(O/H)=8.00  \citep{vanzee06}. 
 For ages $\ge$2 Gyr, we also show the  Z=0.0004 ($\sim$1/40 solar) metallicity models.  c): same as b), but  with no models superimposed. d) CMD of stars in an adjacent external field at 4\arcmin$<$SMA$<$6\arcmin  containing approximately the same number of stars as b), and mostly sampling background galaxies and MW foreground stars.}
 \label{u685_cmd}
\end{figure*}

\begin{figure*}
\includegraphics[width=\textwidth]{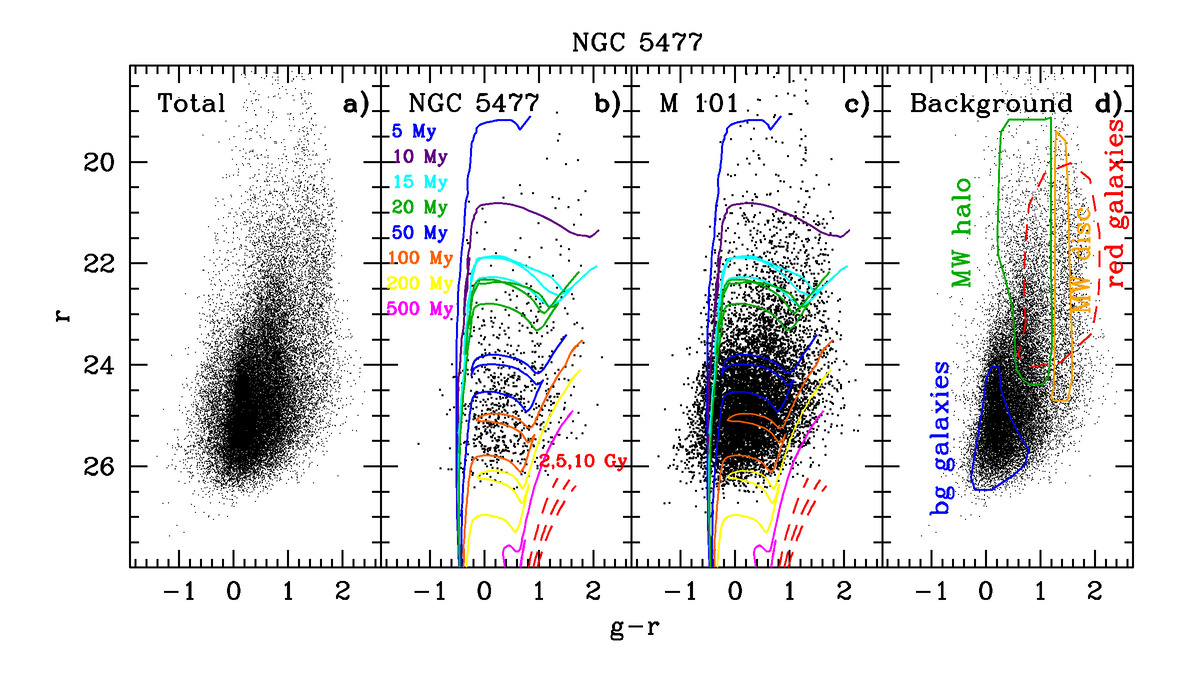}
    \caption{$r$, $g-r$ CMD of NGC~5477. a): total CMD over the whole 23\arcmin$\times$23\arcmin LBC field of view. 
    b): CMD of stars within an elliptical region of semi-major axis (SMA) = 2 \arcmin centered on NGC~5477; overplotted are the PARSEC isochrones shifted to a distance of D=6.8 Mpc [(m-M)$_0$=29.2] and corrected for a foreground reddening of E(B-V)=0.01. The isochrones' metallicity of Z=0.002 is consistent with the H~II region oxygen abundance of 12+log(O/H)=7.9 \citep{berg12}.  c): CMD of a region including the spiral galaxy M~101, according to the contour displayed in Fig.~\ref{five_gals}. 
    d): CMD of stars outside the regions including NGC~5477 and M~101, mostly sampling background galaxies, and foreground MW halo and disc stars, as labelled in the plot. }
    \label{cmd_n5477}
\end{figure*}

\begin{figure*}
\includegraphics[width=\textwidth]{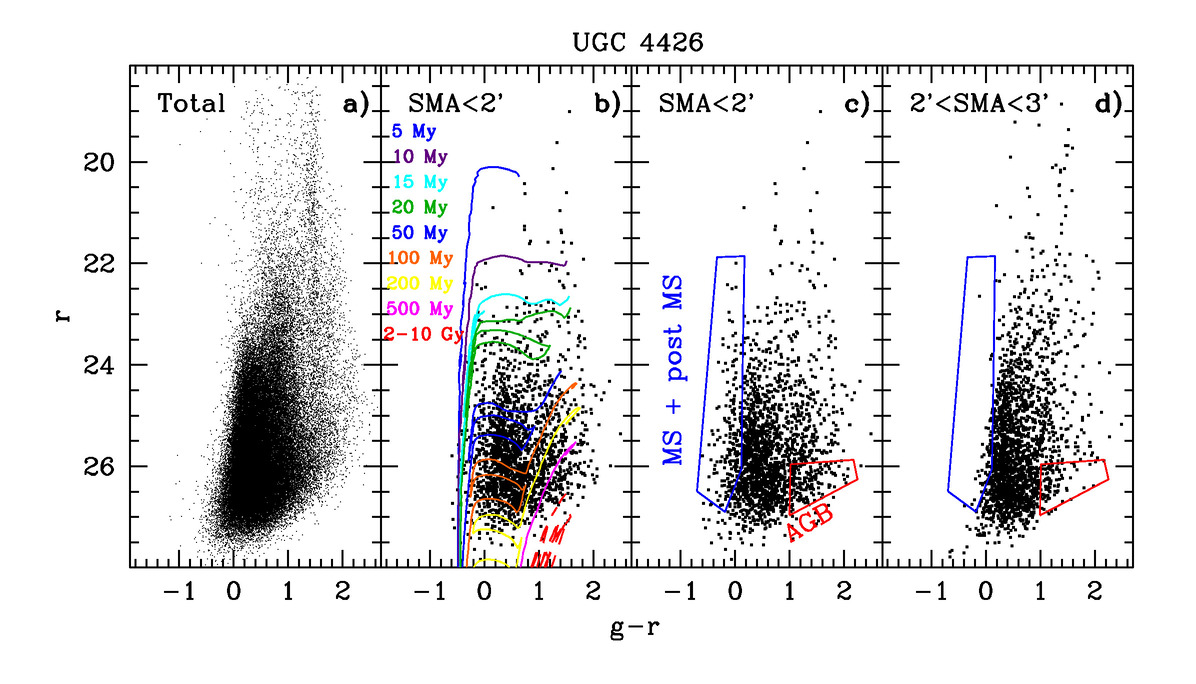}
    \caption{$r$, $g-r$ CMD of UGC~4426. a): total CMD over the whole 23\arcmin$\times$23\arcmin LBC field of view. 
    b): CMD of stars within an elliptical region of semi-major axis (SMA) = 2\arcmin ($\sim$6 kpc) centered on UGC~4426, with overplotted  the PARSEC isochrones shifted to a distance of D=10.3 Mpc [(m-M)$_0$=30.1] and corrected for a foreground reddening of E(B-V)=0.03. The isochrones'  metallicity of Z=0.0007 is consistent with the H~II region oxygen abundance of 12+log(O/H)=7.4 \citep{Pustilnik16}.  c): same as b), but with no models superimposed. The regions sampling MS plus post MS stars and AGB stars are indicated on the CMD. d) CMD of stars in an adjacent external field at 2'$<$SMA$<$3' containing approximately the same number of stars as in b).}
    \label{cmd_u4426}
\end{figure*}

\subsection{NGC~2366}

The LBT/LBC image of NGC~2366, at a distance of D=3.2 Mpc, is shown in panel b) of Fig.~\ref{five_gals}, while its $r$, $g-r$ CMD is displayed in Fig.~\ref{cmd_n2366}.
As for the case of UGC~12613, we show both the total CMD, obtained from all sources photometred in the LBC field of view, and two separate CMDs corresponding to sources located inside and outside an elliptical region of semi-major axis (SMA) = 10\arcmin. Stellar evolutionary features associated to the dwarf galaxy are not identified in the external field CMD, which mostly samples foreground MW stars and unresolved background galaxies, and looks exactly like that 
in the outer field of view around UGC~12613 (Fig.~\ref{cmd_u12613}). 

 On the other hand, we recognize in the CMD at  SMA$<$10\arcmin~ 
the typical evolutionary sequences associated to young, intermediate-age, and old stellar populations. For an easier visualization, the main stellar evolution phases have been indicated in panel c) of Fig.~\ref{cmd_n2366}, while the PARSEC isochrones, shifted to a distance of D=3.2 Mpc [(m-M)$_0$=27.5] and corrected for a foreground reddening of E(B-V)=0.03, have been superimposed on the observed CMD in panel b). 
The plotted isochrones have ages from 10 Myr to 10 Gyr and metallicity Z=0.002, consistent with the $\sim$1/8 solar metallicity derived in the giant H~II region complex Mrk~71 
\citep[i.e., 12 + log(O/H)$\sim$7.9$-$8.0;][]{Izotov97,Noeske00,James16}. 
The vertical blue plume at 
$g-r\sim-0.2$, $r\gsim 21$ encloses stars younger than 200 Myr both in the main-sequence (MS) phase and at the hot edge of the core helium burning - blue loop - phase (post-MS). 
The red vertical plume at $g-r\sim1.4$, $r\gsim19$ is populated by young red supergiants (RSGs) at the brighter magnitudes, and by intermediate-age 
asymptotic giant branch (AGB) stars at fainter luminosities, 
although contamination from foreground MW stars can be significant here. 
Sources of intermediate colours,  $g-r\sim0.5$, and fainter than $r\gsim 24$, are consistent with blue-loops of intermediate-mass stars, with ages from $\sim$500 Myr to $\sim$100 Myr, but a major contamination from background galaxies occurs in this region of the CMD. 
Finally, the concentration of stars at $g-r\sim1$, $r\gsim24.5$ is the RGB of low-mass, old (age$>$1-2 Gyr) stars. This RGB is shallower and much broader in color than that observed in the closer Pegasus dwarf, reaching only down to $\sim$1.5 mag below the tip. This prevents a robust constraint on the stellar metallicity, although it is reasonable to exclude from panel b) of Fig.~\ref{cmd_n2366} a metallicity as high as Z=0.002. 

In order to characterize the spatial behaviour of the resolved stars in NGC~2366, we displayed in Fig.~\ref{n2366_cmd_radial} the CMDs for stars selected 
in elliptical annuli of increasing semi-major axes.
RGB stars can be identified in the CMDs out to SMA$\sim$10\arcmin  ($\sim$9.3 kpc), while at larger galacto-centric distances we only recognize features due to MW foreground stars and background galaxies. The lack of RGB stars in the internal SMA$<$5\arcmin region is not real, but is due to the severe crowding hampering the detection of faint stars. On the other hand, the young population is concentrated toward the internal galaxy regions, with stars  younger than 
$\sim$50 Myr confined within SMA$\lesssim$4\arcmin. This result is consistent with what is typically found for dIrrs and BCDs \citep[see, e.g.,][]{Tosi01,Annibali13,Cignoni13,Sacchi16}. We also notice that the Z=0.002 models, while providing a satisfactory agreement with the CMD of the young population, are always redder than the bulk of the observed RGB stars. This suggests that stars older than 1$-$2 Gyr are significantly metal-poorer than Z=0.002. In particular, a metallicity as low as  Z=0.0004 ($\sim$1/40 solar) 
appears consistent with both the RGB color and slope. We are thus witnessing the presence of an age-metallicity relation in NGC~2366, as usually found in galaxies, but not an evident metallicity gradient in stars older than 1-2 Gyr.

\subsection{UGC~685}

The LBC image of UGC~685, at a distance D=4.7 Mpc, is displayed in panel c) of Fig.~\ref{five_gals}, while its CMD is shown in Fig.~\ref{u685_cmd}. This galaxy lies at the largest distance where we expect to resolve individual RGB stars from the ground in seeing-limited mode, and indeed the detection of the RGB on the displayed total CMDs in panel a) may be challenging. Nevertheless, panel b) and c) of Fig.~\ref{u685_cmd}, showing the CMD within an elliptical region of SMA=4\arcmin  ($\sim$5.4 kpc), demonstrate that the RGB is detected in this case, thanks to a seeing as good as $\sim$0.7\arcsec. The detection appears quite convincing once we compare the internal CMD with an adjacent external field at 4\arcmin$<$SMA$<$6\arcmin  containing approximately the same number of stars:  from the comparison of panel c) and d), one can immediately see that a concentration of stars at r$\gsim$25, $1\lesssim g-r\lesssim1.8$, compatible with the locus of the RGB models, is present in the internal field but not in the external adjacent region, and thus belongs to the galaxy. 

We also notice that no young blue stars are present in the CMD of UGC~685, likely because severely blended to other sources and therefore removed by our selection criteria. 
 From HST CMDs it is however evident that a young population is present in the innermost regions of this galaxy \citep{Sabbi18}.  \cite{Cignoni19} derived its SFH 
 through UV data from the HST Treasury program LEGUS \citep[Legacy Extra-Galactic UV Survey,][]{Calzetti15} and showed that the galaxy has been quite actively forming stars during the last $\sim$200 Myr.

 \subsection{NGC~5477}

NGC~5477, at a distance of D=6.8 Mpc and fairly close to M101, is shown in panel d) of Fig.~\ref{five_gals}, while its CMD is displayed in Fig.~\ref{cmd_n5477}. The CMD, as explained in section~\ref{section_sel_cuts}, results from less stringent selection cuts than in the case of UGC~12613, NGC~2366 and UGC~685, since the photometry was shallower, because of poorer seeing conditions ($1.1\arcsec)$, and we are in a situation where severe crowding prevents to resolve individual stars; therefore  we accept to retain bright sources even if blended to other objects. Blends of two or more stars can still be useful to trace the galaxy stellar population. 

We notice that in the total CMD displayed in panel a), the MW disk sequence appears less defined than in Figs.~\ref{cmd_u12613},\ref{cmd_n2366},\ref{u685_cmd}; this is due to the contamination from bright and extended red sequence galaxies, efficiently removed in the previous examples by the selections in $g-g_{ap}$, $r-r_{ap}$  and $r_h(g,r)$, but retained in the photometric catalog of NGC~5477. Panel b) and c) of Fig.~\ref{cmd_n5477} show the CMDs for two regions centered on NGC~5477 and on M~101, according to the elliptical contours displayed in Fig.~\ref{five_gals}. A comparison with stellar isochrones shifted to the galaxies' distance shows that we are far from reaching the RGB, with the tip expected at r$>$26; 
on the other hand, both CMDs show a population of blue plume stars (MS and post-MS stars) at $g-r\sim-0.4$, $21.5\lesssim r \lesssim 25.5$,  with ages $\leq$50 Myr. The presence of these stars in the photometric catalog is particularly striking once the total CMD in panel a) is compared with the CMD in panel d) obtained from all sources  
outside the regions including NGC~5477 and M~101.

\subsection{UGC~4426}

UGC~4426, at D=10.3, is the most distant target in our sample. Its LBC image is shown in panel e) of Fig.~\ref{five_gals}, and its CMD in Fig.~\ref{cmd_u4426}. The total CMD in panel a) is dominated by background galaxies and MW stars, and no noticeable feature associated to UGC~4426 is visible. Panels b) and c) display the CMD for an elliptical region of SMA$\leq$2\arcmin centered on UGC~4426, while panel d) displays the  CMD for an adjacent outer field at 2\arcmin$<$SMA$<$3\arcmin containing approximately the same number of stars as the internal field. 
Stellar isochrones, superimposed to the CMD in panel b), show that the RGB tip is expected at r$\sim$27, about 1 mag below the faintest stars resolved with our photometry. 
A comparison between panel c) and d) reveals in UGC~4426 an excess of blue stars at $g-r\sim-0.2$, $23 \lesssim r \lesssim 26$ compared to the background field; these  objects fall in the locus of young (age$<$100 Myr) blue MS and post-MS stars. There is also a concentration of red stars at  $g-r>1$, $r>26$ consistent with AGB stars with ages$>$500 Myr, but we can not exclude that these objects are blends of fainter RGB stars.

\section{Stellar Density Maps} \label{section_starmap} 

The CMDs presented in section~\ref{section_cmd} provide a powerful diagnostic to separate stars belonging to the target galaxies from background and foreground contaminants. 
This technique has been extensively used in the literature to identify low surface brightness stellar streams, satellites, and extended stellar halos in galaxies within the Local Group and beyond
\citep[e.g.,][]{Belokurov06,Bernard12,Bonaca12,Beccari14,Greggio14,Ibata14,Martin14,Roderick15,Belokurov16,pisces,Higgs16,Mackey16,Tanaka17,McConnachie18,Pucha19}. 
This approach allows to typically reach much fainter surface brightness structures than from the integrated light, down to values as low as 31-35 mag arcsec$^{-2}$, depending on the faintest 
evolutionary phase sampled by the CMD.

\begin{figure*}
\includegraphics[width=\textwidth]{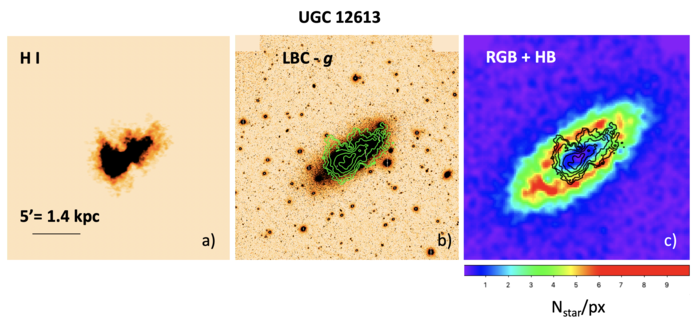}
    \caption{{\bf Panel a):} H~I emission image of UGC~12613 from  LITTLE THINGS \citep{little_things}. {\bf Panel b):} LBC image in $g$ with superimposed the H~I contours for mass densities of  $\sim$0.4, 0.8, 1.6, and 3.2 M$_{\odot}$ pc$^{-2}$ \citep[see][]{iorio17}. {\bf Panel c):} density map of RGB plus HB/RC stars in number of stars per pixel (where the pixel size is $\sim10\arcsec\times10\arcsec$) selected from the CMD of UGC~12613 (see Fig.~\ref{cmd_u12613}); superimposed are the H~I contours. All three images are displayed on the same spatial scale, with a field of view of 
    $\sim23.5\arcmin\times23.5\arcmin$.}
    \label{u12613_sm}
\end{figure*}

\begin{figure*}
\includegraphics[width=\textwidth]{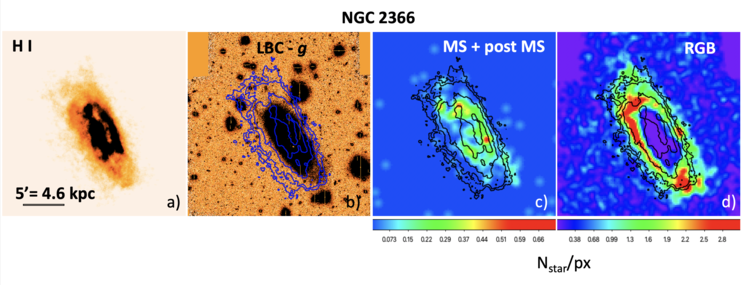}
 \caption{{\bf Panel a):} H~I emission image of NGC~2366 from  LITTLE THINGS \citep{little_things}. {\bf Panel b):} LBC image in $g$ with superimposed the H~I contours for mass densities of  $\sim$0.6, 1.8, 5.4, and 16.2 M$_{\odot}$ pc$^{-2}$ \citep[see][]{iorio17}. {\bf Panel c):} 
  density map of blue MS and post MS stars with age $\lesssim$200 Myr selected within the box outlined in Fig.~\ref{cmd_n2366}; the scale is in number of 
  stars per pixel, where the pixel size is $\sim10\arcsec\times10\arcsec$. Superimposed are the H~I contours.  {\bf Panel d):} Same as c) but for RGB stars, with ages$>$1-2 Gyr . 
 All four images are displayed on the same spatial scale, with a field of view of $\sim23\arcmin\times23\arcmin$.}
 \label{n2366_sm}
\end{figure*}

\begin{figure*}
\includegraphics[width=\textwidth]{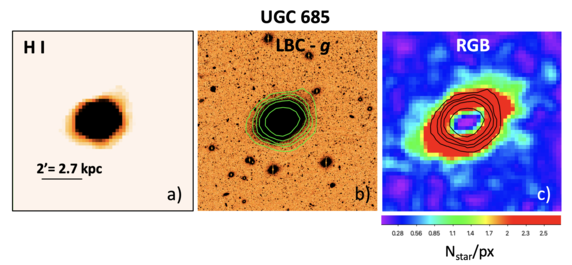}
 \caption{{\bf Panel a):} H~I emission image of UGC~685 from the FIGGS survey \citep{Begum08}, kindly provided by S. 
  Roychowdhury. {\bf Panel b):} LBC image in $g$ with superimposed the H~I contours for mass densities of  $\sim$0.1, 0.5, 1, 2, and 5 M$_{\odot}$ pc$^{-2}$. {\bf Panel c):} Density map of RGB stars in counts per pixel  (pixel size $=10\arcsec\times10\arcsec$) with superimposed the H~I contours. All four images are displayed on the same spatial scale, with a field of view of $\sim9\arcmin\times9\arcmin$.
  }
 \label{u685_sm}
\end{figure*}

\begin{figure*}
\includegraphics[width=\textwidth]{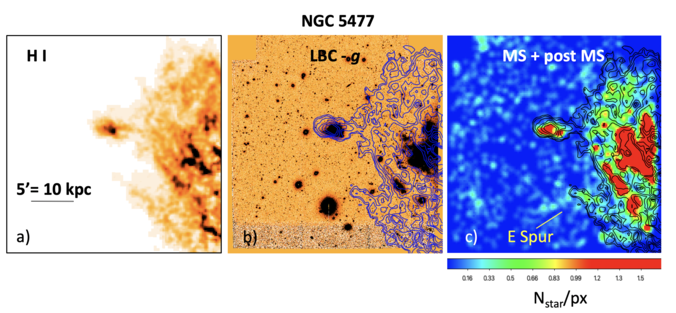}
 \caption{{\bf Panel a):} H~I emission image of NGC~5477 from  WHISP \citep{Swaters02}. {\bf Panel b):} LBC image in $g$ with superimposed the H~I contours for mass 
 densities of $\sim$1, 3, 4, 5, 8 and 11 M$_{\odot}$ pc$^{-2}$.
 {\bf Panel c):} density map of young (age $\lesssim$100 Myr) MS and post MS stars in number of stars per pixel 
 (pixel size = $10\arcsec\times10\arcsec$) selected from the CMD of NGC~5477 (see Fig.~\ref{cmd_n5477}); superimposed are the H~I contours. On the density map, we indicate the E Spur 
  identified by \citet{Mihos13} from deep optical images. All three images are displayed on the same spatial scale, with a field of view of $\sim25\arcmin\times24\arcmin$. 
 }
  \label{n5477_sm}
\end{figure*}

\begin{figure*}
\includegraphics[width=\textwidth]{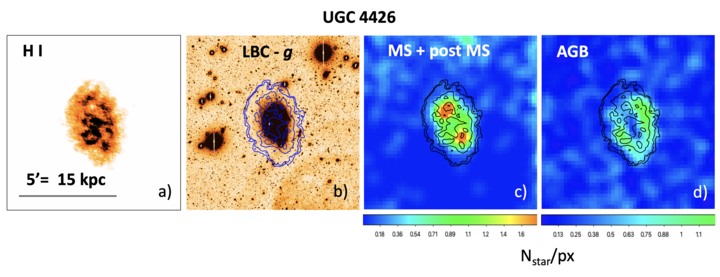}
 \caption{{\bf Panel a):} H~I emission image of UGC~4426 from  LITTLE THINGS \citep{little_things}. {\bf Panel b):} LBC image in $g$ with superimposed the H~I contours for mass densities of  $\sim$0.5, 1, 3, and 5 M$_{\odot}$ pc$^{-2}$. {\bf Panel c):} density map of young (age $\lesssim$100 Myr) MS and post MS stars in number of stars per pixel (where the pixel size is  $10\arcsec\times10\arcsec$) selected from the CMD of UGC~4426 (see Fig.~\ref{cmd_u4426}); superimposed are the H~I contours. 
{\bf Panel d):} density map of sources with $g-r\gtrsim1$, $r\gtrsim26$, which are AGB stars with age$>$500 Myr or blends of two or more fainter RGB stars. 
  All three images are displayed on the same spatial scale, with a field of view of $\sim9\arcmin\times9\arcmin$.}
\label{u4426_sm}
\end{figure*}

In Figs.~\ref{u12613_sm} through ~\ref{u4426_sm} we present stellar density maps for UGC~12613, NGC~2366, UGC~685, NGC~5477, and UGC~4426. The maps are based on RGB and HB/RC stars down to r$\sim$26 for UGC~12613, and on RGB stars down to r$\sim$26.3 and  r$\sim$26.5  for NGC~2366 and UGC~685, respectively, corresponding to limits of $\sim$2 mag and $\sim$1.5 mag  below the RGB tip. The density maps of NGC~5477 and UGC~4426 were instead constructed from evolutionary phases younger than the HB/RC or RGB, since the galaxies' distances prevent to reach even the latter.  For NGC~5477, the stellar density map traces blue MS and post-MS stars, with ages$\lesssim$100 Myr. For UGC~4426, the photometry is quite deeper than usual (reaching down to r$\sim$27) and allowed us to use, besides  blue MS and post-MS stars, also AGB stars at magnitudes just brighter than the expected RGB tip location, with ages$>$500 Myr. 
The displayed maps provide the density of truly observed star counts, not corrected for incompleteness; as a consequence, the maps of RGB and HB/RC stars exhibit a ``hole'' in the central, more crowded galaxy regions, due to loss of stars counts in the increasingly unresolved stellar background. On the other hand, incompleteness is less important in the external galaxy regions, allowing us to trace the outer low surface brightness component.

The stellar maps were compared with H~I data from different sources: the Local Irregulars That Trace Luminosity Extremes, The H~I Nearby Galaxy Survey \citep[LITTLE THINGS,][]{little_things} 
for UGC~12613, NGC~2366, and UGC~4426; the Westerbork observations of neutral Hydrogen in Irregular and SPiral galaxies  \citep[WHISP,][]{Swaters02} for NGC~5477; and  
the Faint Irregular Galaxies GMRT Survey \citep[FIGGS,][]{Begum08,Roy09} for UGC~685. 
For the LITTLE THINGS data, we adopted the naturally weighted  total intensity maps \footnote{https://science.nrao.edu/science/surveys/littlethings/data}, that are more sensitive to large scale structures than the robust weighted maps. The WHISP data at a 30\arcsec$\times$30\arcsec resolution were obtained from the  ASTRON http://wow.astron.nl web site. The high sensitivity, low resolution ($42\arcsec\times40\arcsec$) FIGGS data of UGC~685 were kindly provided to us by S. Roychowdhury \citep[see also][for higher resolution data of this galaxy]{Roy09}. 
For each galaxy, we show the H~I emission image and the H~I contours superimposed both to the LBC $g$ image and to the stellar density maps. To produce H~I contours in units of M$_{\odot}$/pc$^2$, we used Eq.(3) of \cite{iorio17}.

Figs.~\ref{u12613_sm},~\ref{n2366_sm} and ~\ref{u685_sm} show that, whenever the CMD is deep enough to sample RGB or HB/RC stars, the old  (age$>$1-2 Gyr) 
stellar population appears extended at least as much as the H~I gas, and possibly even more; the case of UGC~12613 is extreme in this sense, with the old stellar component, detected down to 
 a surface brightness limit of $\sim$31 mag arcsec$^{-2}$ (see section~\ref{section_profiles}), 
extending  in projection as far as twice the faintest H~I contour at  
$\sim$0.4 M$_{\odot}$ pc$^{-2}$ \citep[see also][]{Mc07,Kniazev09}. Here the difference between the star count 
and the H~I extensions is striking, and we investigated whether it could be due to a limited sensitivity of the 
LITTLE THINGS, high resolution H~I data: a comparison of the H~I contours displayed in Fig.~\ref{u12613_sm} with higher sensitivity VLA H~I data from \citet{Young03} and with very deep  Arecibo Legacy Fast ALFA extragalactic H~I survey (ALFALFA) data \citep{alfalfa,Kniazev09} reveals comparable H~I extensions, indicating that the neutral gas is indeed much less extended than the stellar component.

Young stars are observed to be less extended than the old stellar component and than the H~I; this is well shown in the stellar maps obtained for NGC~2366 in Fig.~\ref{n2366_sm}, where stars younger than $\sim$200 Myr 
are  more centrally concentrated than the old stellar component, and do not reach as far as the H~I extension. This is a typical behavior of the stellar populations in star-forming dwarfs, where the youngest stellar populations preferentially clump toward the most central galaxy regions, while old stars are homogeneously distributed and extend out to large galacto-centric distances \citep[see][for a review]{Tolstoy09}. 

In Figs.~\ref{n5477_sm} and \ref{u4426_sm}, the density maps for NGC~5477 and UGC~4426 provide partial knowledge of the existing stellar populations, since the CMD is far from reaching the locus of old RGB stars. Nevertheless, it is possible to gather some information from the distribution of young blue MS and post MS stars and of intermediate-age AGB stars. 
The density map for  the dwarf NGC~5477 and for a portion of the spiral M~101 in Fig.~\ref{n5477_sm} was derived from stars at 21$\lesssim r\lesssim$26, $g-r\lesssim-0.1$, with ages$\lesssim$100 Myr (see Fig.~\ref{cmd_n5477}). Young star counts trace quite well the H~I distribution. For M~101, we recover the E spur, a low surface brightness feature extending from M~101 to East, previously identified by \cite{Mihos13} from deep optical images; this stellar feature spatially corresponds to a similar morphology feature observed in the H~I.  The H~I emission map also shows a ``bridge'' connecting NGC~5477 to M~101, however this has no visible counterpart in our stellar density map.

As for UGC~4426, we provide in Fig.~\ref{u4426_sm} maps both for the young (age$\lesssim$100 Myr) stellar component and for an older population composed of AGB stars with age$>$500 Myr plus possible blends of older RGB stars (see the regions outlined on the CMD of Fig.~\ref{cmd_u4426} for the adopted selection).  
A comparison of the stellar density maps with the $g$-LBC and H~I emission images indicates that, while young stars agree with the galaxy extension outlined by the optical image, 
the older AGB component extends further away, reaching out to the faintest H~I contour at $\sim$0.5 M$_{\odot}$/pc$^2$. This indicates that, even for the most distant targets where individual RGB stars can not be resolved from the ground, stellar photometry is a valuable tool for tracing the galaxy stellar population down to fainter surface brightness sensitivities than allowed by the integrated light.

\section{Surface Brightness Profiles} \label{section_profiles}

\begin{figure}
\includegraphics[width=\columnwidth]{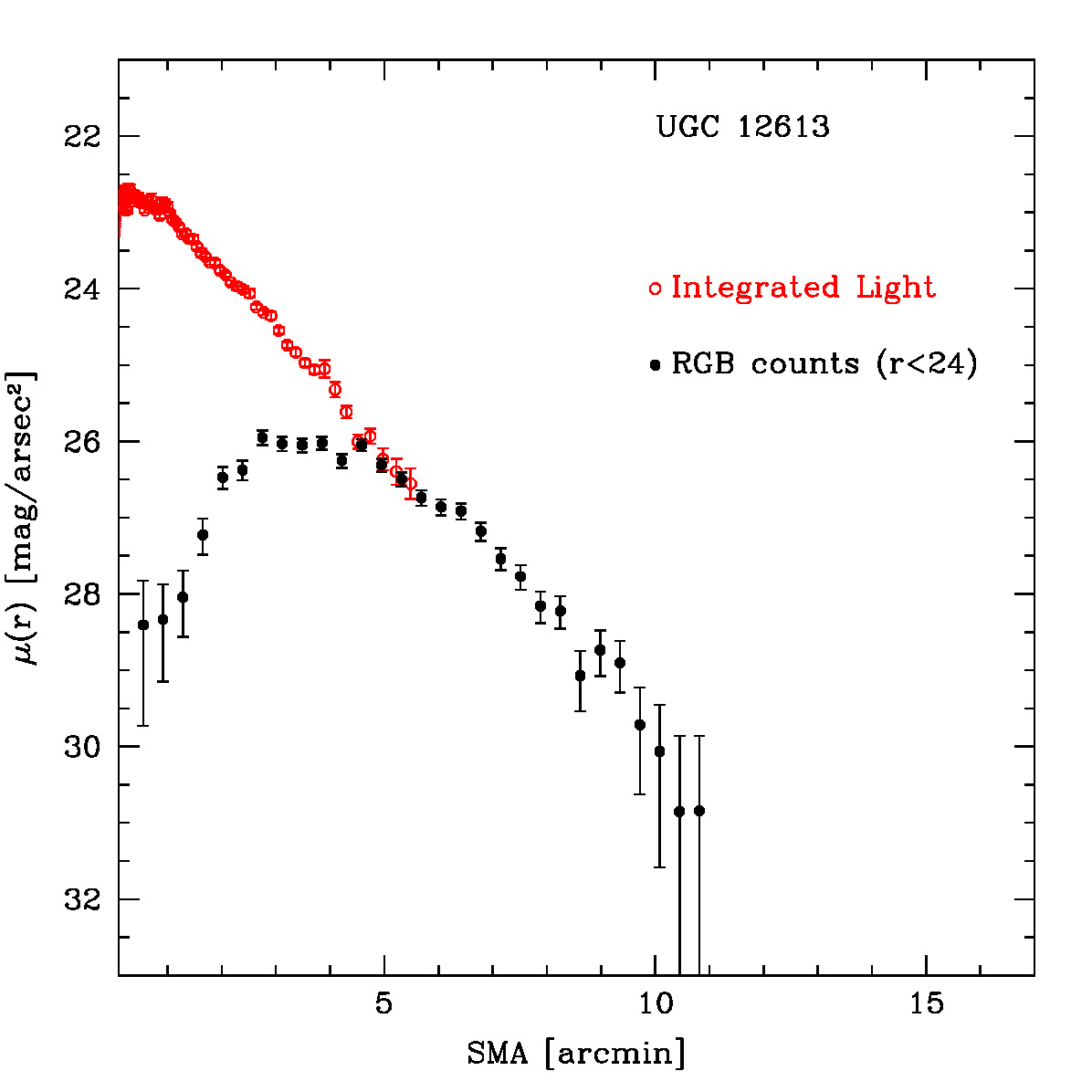}
    \caption{Surface brightness profile $\mu(r)$ in the r band for UGC~12613 (Pegasus dwarf irregular). The profile was calculated in elliptical annuli of increasing semi-major axis (SMA). Red open symbols denote the integrated light profile obtained through the IRAF {\texttt ellipse} task, while full symbols correspond to the profile from RGB star counts brighter than r=24. A vertical shift was applied to the RGB count profile to match it to the integrated light profile at 4.5\arcmin$\lesssim$SMA$\lesssim$6\arcmin. Both integrated-light and star-count profiles are background subtracted (see section~\ref{section_profiles} for details). 
  }
    \label{fig_profile}
\end{figure}

Integrated light profiles in $g$ and $r$ were derived through the {\texttt ellipse} task in IRAF, which is based on the iterative  method  described  by \cite{Jed87}. 
The light intensity as a function of  the semi-major axis (SMA)  was computed as the mean flux  within concentric elliptical annuli of fixed ellipticity ($\epsilon=1-b/a$) and position angle (PA),  where the ellipse parameters were derived iteratively by fitting isophotes to the galaxy image in the $r$ band. 
Bright foreground stars and obvious bright background galaxies were masked out; 
however, to account for the remaining fainter population of background galaxies, we subtracted to the derived profile a constant background level computed as the average surface brightness in the external regions where the profile flattens out.  The final errors were calculated by summing the intensity uncertainty provided by the ellipse task and the standard deviation of the background in quadrature. As an example, we show in Fig.~\ref{fig_profile} the $r$ band profile computed for UGC~12613  in elliptical annuli 
with ellipticity $\epsilon=0.55$ and major axis position angle PA$=125^{\circ}$; we consider the integrated-light profile robust out to SMA$\sim6\arcmin$, while 
we observe very large errors, mainly due to the dominant effect of the background uncertainty, at larger distances.

Resolved star counts allow us to reach much outer galaxy regions than the integrated light; for UGC~12613, for instance,  the stellar density map displayed in Fig.~\ref{u12613_sm} shows that stars are observed out to SMA$\sim$12\arcmin  ($\sim$3 kpc at the galaxy's distance of $\sim$0.97 Mpc). RGB stars trace the distribution of the stellar populations older than 1-2 Gyr, and are typically considered good tracers of the stellar mass \citep[see e.g.][]{rys11}. In Fig.~\ref{fig_profile},  we show the RGB count profile computed for UGC~12613, after having applied a vertical shift to match the integrated light profile at  4.5\arcmin$\lesssim$SMA$\lesssim$6\arcmin where the two profiles overlap with the same slope. To calculate the profile, we selected RGB stars brighter than $r=24$ in order to minimize incompleteness effects toward the most central, crowded galaxy regions. Indeed, the CMDs displayed in Fig.~\ref{u12613_cmd_radial} for annuli of increasing galacto-centric distances 
show that the UGC~12613 photometry reaches down to r$\sim$27 at SMA$\geq$4\arcmin, and therefore it is reasonable to assume a $\sim$100\% completeness for counts brighter than $r=24$ at these galacto-centric distances. 
On the other hand, the RGB counts are highly incomplete within the internal SMA$<$4\arcmin  region, which causes a flattening and drop of the profile toward the galaxy center. 
The star count profile displayed in Fig.~\ref{fig_profile} was background-subtracted; in fact, to account for the contamination from unresolved background galaxies to the RGB counts,  we subtracted to the profile a constant level computed from the counts at SMA$>$12\arcmin, where the stellar density map shows no significant signal from the galaxy stellar population. 

The combined integrated light/ RGB count profile indicates that we reach a surface brightness sensitivity at least as faint as $\mu(r)\sim$ 31 mag arcsec$^{-2}$  for UGC~12613; indeed, fainter HB/RC stars may allow for even deeper surface brightness limits in the most external regions. 
In subsequent papers where we will present a dedicated study of the surface brightness profiles 
for the galaxies in the SSH sample.

\subsection{Sensitivity for unresolved structures: the example of an extreme UDG} \label{section_udg}

Our deep wide field images capture several interesting objects lying in the background of our target galaxies, like, e.g., interacting galaxies, galaxy clusters and groups, low surface brightness  (SB) galaxies etc. Here we briefly report on the serendipitous discovery of a very low SB galaxy, an Ultra Diffuse Galaxy \citep[UDG, see][]{udg,koda}, in the background of UGC~1281, as it may serve as an example of the sensitivity of the survey to low SB structures not resolved into stars. 

\begin{figure*}
\includegraphics[width=\textwidth]{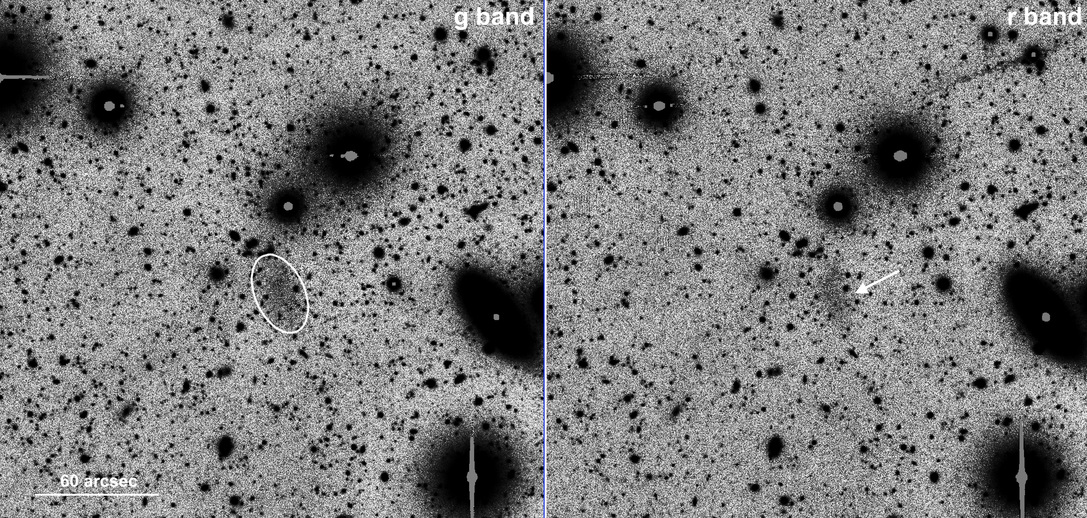}
    \caption{g band (left panel) and r band (right panel) images of the newly discovered Ultra Diffuse Galaxy 
    SSH-UDG~1. The galaxy is highlighted by an ellipse of semi-major axis a=$20.0\arcsec$, axis ratio $b/a=0.6$ and position angle PA=$24.0\degr$ in the left panel, and by an arrow in the right panel. North is up and East is to the left. 
  }
    \label{UDGima}
\end{figure*}

Zoomed images in g and r bands of the newly discovered galaxy, that we dub SSH-UDG~1 (hereafter UDG~1, for brevity), are shown in Fig.~\ref{UDGima}. The system appears as a slightly irregular and fuzzy, elongated light cloud at the center of the images. 

 \begin{table}
  \caption{Main properties of  SSH-UDG~1}
  \label{udg_tab}
  \begin{tabular}{lcc}
\hline
quantity    & value   & units \\
\hline
RA          & 27.46218       & deg \\
Dec         & +32.62831      & deg \\
$r_e$       & 11.3$^a$       & arcsec     \\
E(B-V)      & 0.039           & mag \\
$g_{int,0}$ & $21.06 \pm 0.1$ & mag \\ 
$r_{int,0}$ & $20.77 \pm 0.1$ & mag \\ 
$V_{int,0}$ & $20.89 \pm 0.1^b$ & mag \\
\hline
  \end{tabular}
\begin{tablenotes}
\small
\item{} All the magnitudes are corrected for reddening.
\item{} The subscript ${int,0}$ denotes integrated magnitudes, corrected for extinction.
\item{a} Average of the fits to the r and g band profiles.
\item{b} Obtained using the transformations by Lupton (2005)\footnote{\tt http://www.sdss3.org/dr9/algorithms/sdssUBVRITransform.php\#Lupton2005}.   
\end{tablenotes}
 \end{table}

\begin{figure}
\includegraphics[width=\columnwidth]{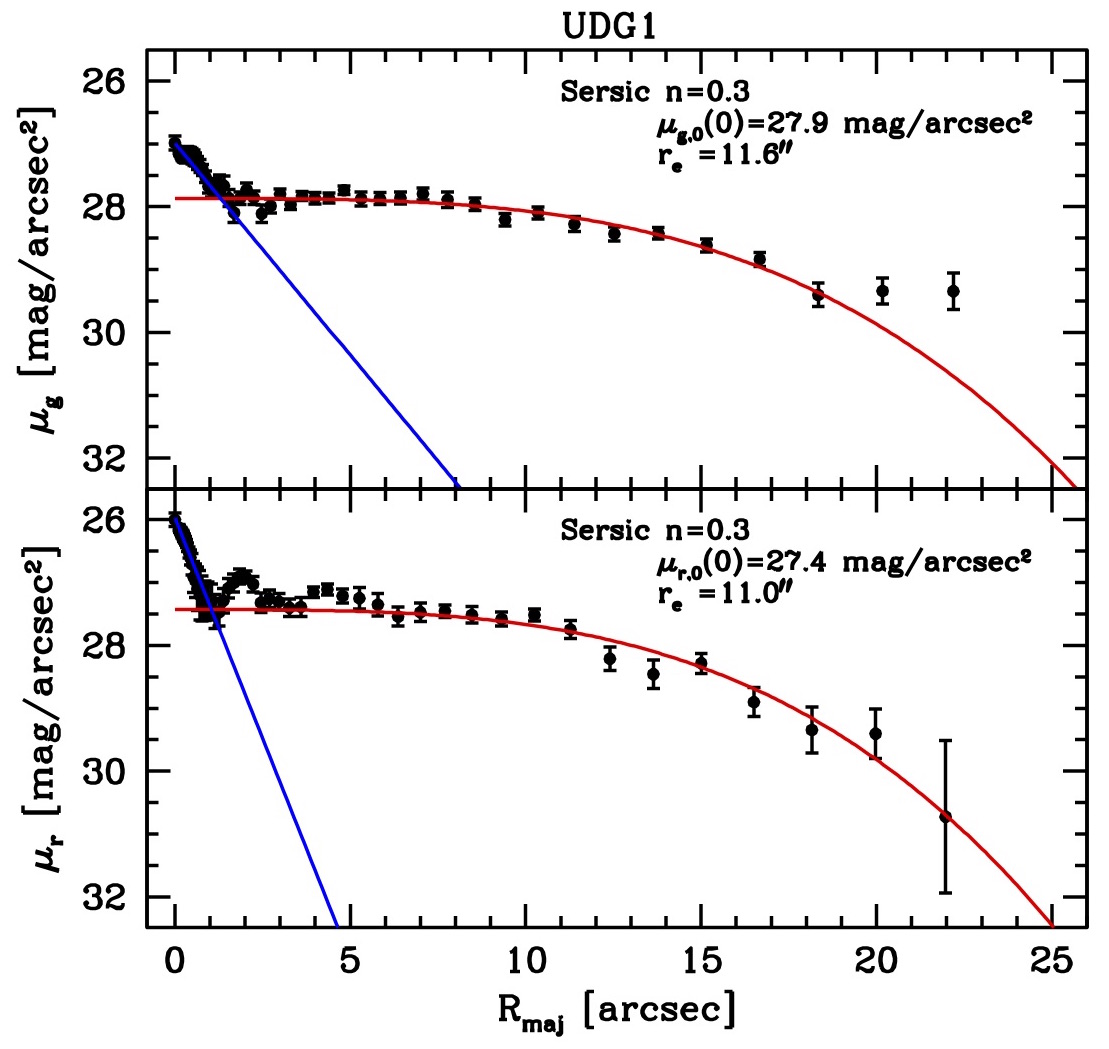}
    \caption{Surface brightness profiles of SSH-UDG~1 in g band (upper panel) and r band (lower panel). The red lines are the best-fit \citet{sersic} models for the body of the galaxy, the blue lines are exponential profiles aimed at reproducing the inner cusp in the profile ($r_e=2.7\arcsec$ and $r_e=1.3\arcsec$ for the g and r band profiles, respectively.).
  }
    \label{UDGprof}
\end{figure}

We used Galfit \citep{galfit} to estimate the position of the center, the axis ratio (${\frac{a}{b}}\simeq 0.6$) and the position angle (PA$=24\degr$). Then, assuming these values,  we derived the SB profile in both passbands with IRAF/ellipse. The two profiles are shown in Fig.~\ref{UDGprof} while the main properties of the system are listed in Tab.~\ref{udg_tab}. The main body of the galaxy is well fitted by a \citet{sersic} profile with 
S\'ersic index lower than unity (n=0.3), as typical for UDGs \citep{koda}, and an extremely low central surface brightness ($\mu(0)>27.0$~mag/arcsec$^2$ in both bands). A central cusp, fitted with an exponential profile in Fig.~\ref{UDGprof}, suggests the possible presence of a central stellar nucleus, that is not unusual in this kind of galaxies \citep{koda}. The integrated color is typical of blue UDGs, that are more frequently found in low density environments \citep[e.g., field, groups][]{rt17,Bellazzini17}. 
Fig.~\ref{UDGprof} shows that we are able to reach a surface brightness as faint as $\simeq 31.0$~mag/arcsec$^2$ not only by star counts in the outskirts of (relatively) nearby galaxies, as shown above, but also tracing diffuse unresolved light in more distant systems. 

UDG~1 is projected unto a substructure of many tens of galaxies with $0.03\le z\le 0.04$, likely associated with the UGC~1306 group. Since UDGs are preferentially found in clusters and groups of galaxies, it is worth considering the hypothesis that also UDG~1 is a member of this group. \citet{Tully13} provide an estimate of the distance to UGC~1306 itself of $D=81.3^{+16.4}_{-13.7}$~Mpc, using the Tully-Fisher relation. Adopting the same distance for UDG~1 we would obtain $M_V\simeq -13.7$ and $r_e\simeq 4.4$~kpc. On the other hand, if we assume the redshift of the likely group member that is the nearest, in projection, to UDG~1 (2MASXI J0149427+323730,  z=0.038361, according to SIMBAD\footnote{\tt http://simbad.u-strasbg.fr/simbad/}), and 
$H_0=75.0$~km~s$^{-1}$~Mpc$^{-1}$, we obtain D=153~Mpc. At this distance UDG~1 would have $M_V\simeq -15.0$ and $r_e\simeq 8.4$~kpc. In any case UDG~1 appears as extremely diffuse. Its central (and effective) surface brightness and S\'ersic index are at the lower limit of the distribution of known UDGs \citep{koda}. 
Its discovery, just by visual inspection of the images, clearly illustrates the high sensitivity of the SSH images to very low SB galaxies and sub-structures.  

Due to its low surface brightness, UDG~1 remains undetected in existing all-sky or large-scale public surveys that cover its field of view (such as the Sloan Digital Sky Survey, the All Wide-field Infrared Survey Explorer program, and the Canada-France-Hawaii Telescope MegaCam Legacy survey), which are much shallower than SSH.

\section{Discussion} \label{section_discussion}

\begin{figure*}
\includegraphics[width=\textwidth]{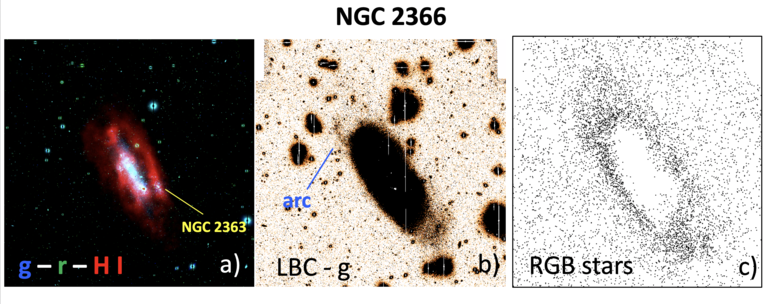}
    \caption{{\bf Panel a:} $g$-$r$-H~I color-combined image of NGC~2366 highlighting the presence of two parallel ridges of H~I that appear to connect NGC~2366 to the giant H~II region complex NGC~2363. {\bf Panel b:} Deep LBC $g$ image with the display cuts chosen to emphasize a low surface brightness arc. {\bf Panel c:} distribution of RGB stars  extending out to much larger galacto-centric distances than visible from the LBC image. All images have been displayed on the same spatial scale. North is up and East is left. 
  }
    \label{fig_n2366_discuss}
\end{figure*}

The stellar density maps derived from RGB and HB/RC counts for UGC~12613, NGC~2366 and UGC~685 in Figs~\ref{u12613_sm},~\ref{n2366_sm}, and ~\ref{u685_sm} provide evidence for extended low surface brightness stellar envelopes around the three dwarf galaxies, reaching as far out as, or even beyond, the observed H~I disk. 
These extended envelopes consist of stars at least 1$-$2 Gyr old, while young stars are  found to be confined toward the innermost galaxy regions, 
 a typical behavior observed in star-forming dwarfs \citep[see][for a review]{Tolstoy09}. 

From a theoretical point of view, the disruption of smaller merging galaxies onto a more massive host has been proposed as a possible mechanism for the origin of extended stellar envelopes. 
According to \cite{Penar06}, the emergence of extended exponential discs is a generic feature of the disruption of satellites on coplanar orbits. Some simulations show that minor merger events are typically responsible for  depositing stellar mass in the galaxy outskirts, whereas relatively massive satellites tend to deposit their stars further inside the host galaxy \citep[e.g.][]{Amorisco17}. According to \cite{Karademir19}, mini-mergers, with mass ratios lower than 1:10, can lead to an increase of the disc size, depending also on the infall direction of the satellite relative to the host disc, without significantly disturbing the center. These mechanisms offer a possible explanation for the 
extended low surface brightness stellar components observed in the SSH dwarfs.

The presence of an extended stellar envelope is particularly striking in the case of UGC~12613 (Pegasus dwarf irregular) where the deep LBT data allow us to trace a low surface brightness ($\mu\gsim$31 mag arcsec$^{-2}$) population of old RGB and RC/HB stars extending out to $\sim$6 times the galaxy half light radius \citep[$r_e=2.1\arcmin$,][]{Mc12}, and largely beyond the observed H~I disk. \citet{Mc07} noticed a ``cometary'' appearance of the H~I as opposed to a regular, elliptical distribution of the stars in Pegasus, and proposed ram pressure stripping as a possible explanation. 

In the case of NGC~2366, support in favor of a merging event comes from the comparison between the H~I emission and the starlight distribution in the innermost regions. 
In fact, the g$-$r$-$H~I color-combined image displayed in panel a) of Fig.~\ref{fig_n2366_discuss} highlights the presence of two ridges of H~I running parallel to the semi-major axis 
that if de-projected appear as a large ring \citep{Hunter01}; the H~I ring connects the central star forming regions of NGC~2366 with the supergiant H~II complex NGC~2363, providing hints that NGC~2363 is a gas-rich satellite being accreted by  NGC~2366 \citep[see also][]{Drissen00}. This scenario is further supported by H~I kinematics:
NGC~2366 has an extended regularly-rotating H~I disk, however a strong kinematic distortion is detected to the North-West \citep{Oh08,Lelli14b,iorio17}, and a steep velocity gradient is observed at the position of NGC~2363, 
suggesting that the NGC~2363/NGC~2366 system is an on-going minor merger \citep{Lelli14b}.
On our deep LBC image, a stellar arc-like feature  is identified $\sim$5 kpc away from the galaxy center (panel b)). Then, faint RGB stars in panel c) trace a low surface-brightness component extending beyond structures visible in the optical image, toward the North-East direction and along the semi-major axis of the NGC~2366 disc; this faint component appears itself structured into arc-like features (see panel d) of Fig.~\ref{n2366_sm} and panel c) of Fig.\ref{fig_n2366_discuss}). We speculate that this extended stellar envelope may have been produced by the accretion of NGC~2363 or of other satellites and a dynamical analysis of the system is currently ongoing.

 \begin{figure}
\includegraphics[width=\columnwidth]{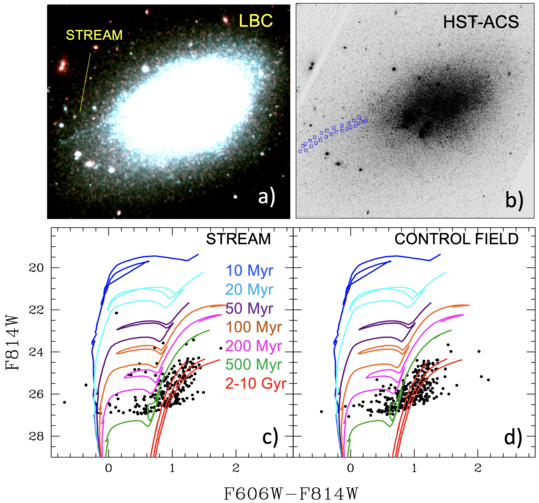}
    \caption{{\bf Panel a):} $g$,$r$ color-combined LBC image of UGC~685 highlighting the thin stream running parallel to the galaxy major axis. The image field of view is $\sim$1.8$\times$1.8 arcmin$^{2}$. {\bf Panel b)}: HST ACS image in F606W, where individual stars are resolved. The image is displayed on the same spatial scale of the LBC image, with identified the region corresponding to the stream. {\bf Bottom panels:} CMDs for the stream (c) and for an adjacent 
    control-field (d) with superimposed the PARSEC stellar isochrones for Z=0.004 and for different ages, as indicated in the legend. The stream 's CMD exhibits a 100$-$200 Myr old blue-loops population not present in the control field. CMDs were constructed from the LEGUS public photometric catalog \citep{Sabbi18}.
 }
    \label{fig_u685_discuss}
\end{figure}

 \begin{figure*}
\includegraphics[width=\textwidth]{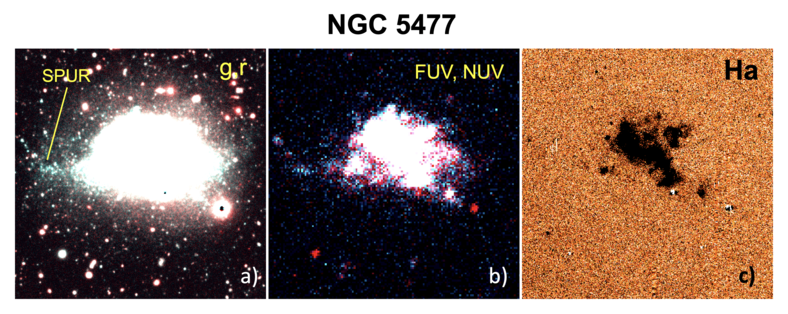}
    \caption{{\bf Panel a):} $g$,$r$ color-combined LBC image of NGC~5477 showing a ``spur'' of blue stars extending from the galaxy toward East.  The image field of view is $\sim$3.5$\times$3.5 arcmin$^{2}$. {\bf Panel b)}: 
On the same scale, GALEX FUV, NUV color-combined image of NGC~5477, where the spur exhibits blue FUV$-$NUV colors. {\bf Panel c):} H$\alpha$ image of NGC~5477 from the Local Volume Legacy (LVL) survey \citep{Dale09}, showing (faint) emission at the spur's position. }
    \label{fig_n5477_discuss}
\end{figure*}

Further interesting features that are worth discussing because potentially associated to interaction, merging, or accretion events are observed for UGC~685 and NGC~5477. Panel a) of Fig.~\ref{fig_u685_discuss} displays our deep LBC image of UGC~685 highlighting a thin ``stream'' of stars running parallel to the galaxy major axis. The stream falls into the field of view of HST ACS data from the LEGUS survey 
\citep{Calzetti15} and is resolved into individual stars there (panel  b) of  Fig.~\ref{fig_u685_discuss}). 
Its recent SFH from the UV data from the LEGUS survey has been derived by \cite{Cignoni19}. 
In order to get insights into its stellar populations, we inspected F606W$-$F814W vs. F814W  CMDs constructed from the photometric catalog of \cite{Sabbi18}, selected for a region containing the stream-like feature and for an adjacent control-field. The derived CMDs are displayed in panels c) and d) of Fig.~\ref{fig_u685_discuss}, and  indicate that both fields are dominated by a population of old ($>$1-2 Gyr) RGB stars; however, we recognize in the stream a  sizeable population of younger blue loop stars with ages of $\sim$100 -200 Myr, not present in the control field. We tested the robustness of this result by checking the absence of this younger population in other similar control-fields, not finding any. We also visually inspected the appearance of these objects on the HST images to exclude that they could be in fact background galaxies, star clusters, or blends of two or more stars. 

Finally, we show in panel a) of Fig.~\ref{fig_n5477_discuss} our deep LBC $g$, $r$ color-combined image of NGC~5477, where we observe a ``spur'' of blue stars departing from the galaxy and extending toward East. The spur is well visible also in GALEX images (panel b)), with rather blue FUV$-$NUV colors compared to the galaxy body, suggesting a very young stellar population. The fact that some H$\alpha$ emission is detected at its position (panel c)) indicates that star formation is still active there. A portion of the spur falls within the LEGUS data of NGC~5477 and is resolved into stars younger than 60 Myr; this confirms a spatial progression of the star formation toward the East direction. 
  
\section{Conclusions} \label{section_conclusions}

SSH is an LBT strategic long-term program imaging 45 late-type dwarf galaxies in $g$ and $r$, with the aim of searching for satellites and/or evidences of accretion/perturbation events down to the smallest mass scales. With SSH we plan to characterise the frequency and the properties of streams and sub-structures around dwarfs outside the Local Group (except for one target, UGC 12613, which is inside), but close enough (namely, closer than 11 Mpc) in the Local Universe to allow us to study their resolved stellar populations with LBT or with HST. The properties of our sample galaxies make SSH complementary to any other existing survey of similar aim we are aware of.

The sample galaxies have been chosen to study the properties of the satellites/streams as a function of the galaxy mass and environment. We expect our database to become a reference set for further studies of these objects, both with follow-up observations (either deeper imaging with more powerful instruments, or ad-hoc spectroscopy) and with hydro-dynamical and N-body modelling of their components.

So far, 40 galaxies in SSH have been observed, while data analysis has been completed for 
about half of the target sample (24 galaxies). In this presentation paper, we describe the goal of the survey, the sample properties, the image acquisition, and our strategies for both the data reduction and the corresponding analyses of the results. We present a few representative cases to illustrate the kind of analyses that we will apply to all the 45 targets  and 
the expected results. In particular, the analysis presented here indicates that we are reaching, for the closets targets which can be resolved into individual stars, a surface brightness limit at least as faint as  $\mu(r)\sim$ 31 mag arcsec$^{-2}$. This sensitivity provides direct 
evidence for the presence of extended low surface brightness stellar envelopes around the dwarfs, reaching farther out than traced by the integrated light. These stellar envelopes extend as far out as, or even beyond, the observed H~I disk. 
Stellar streams, arcs, and peculiar features are detected for some of the dwarfs, indicating possible perturbation, accretion, or merging events.

We also report on the discovery of an extreme case of Ultra Diffuse Galaxy ($\mu_g(0)=27.9$~mag/arcsec$^2$) in the background of one of our targets, to illustrate the power of the survey in revealing extremely low surface brightness systems.

When SSH is completed, we will present its overall results in a concluding paper. Particularly interesting cases will be presented individually.

\section*{Acknowledgements}

We thank the anonymous referee of his/her useful comments and questions, which helped to improve the paper. 

We thank G. Iorio and B. Holwerda for useful discussion. We are deeply indebited to S. Roychowdhury for giving us the FIGGS H~I data of UGC~685.  

This research is partially funded through the INAF Main Stream program SSH 1.05.01.86.28. F. A., M.C. and M.T. acknowledge funding from INAF PRIN-SKA-2017 program 1.05.01.88.04. 
We acknowledge the support from the LBT-Italian Coordination Facility for the execution of observations, data distribution and pre-reduction. 
LBT is an international collaboration among institutions in the United States, Italy and Germany. LBT Corporation partners are: the University of Arizona on behalf of the Arizona Board of Regents; Istituto Nazionale di Astrofisica, Italy; LBT Beteiligungsgesellschaft, Germany, representing the Max-Planck Society, the Leibniz Institute for Astrophysics Potsdam, and Heidelberg University; the Ohio State University, and the Research Corporation, on behalf of the University of Notre Dame, University of Minnesota and University of Virginia.

We have made use of the WSRT on the Web Archive. The Westerbork Synthesis Radio Telescope is operated by the Netherlands Institute for 
Radio Astronomy ASTRON, with support of NWO. 
The WHISP observation were carried out with the Westerbork Synthesis Radio Telescope, which is operated by the Netherlands Foundation for Research in Astronomy (ASTRON) with financial support from the Netherlands Foundation for Scientific Research (NWO). The WHISP project was carried out at the Kapteyn Astronomical Institute by J. Kamphuis, D. Sijbring and Y. Tang under the supervision of T.S. van Albada, J.M. van der Hulst and R. Sancisi.

This research has made use of the SIMBAD database,
operated at CDS, Strasbourg, France 

This research has made use of the NASA/IPAC Extragalactic Database (NED),
which is operated by the Jet Propulsion Laboratory, California Institute of Technology,
under contract with the National Aeronautics and Space Administration.

Part of the analysis presented in this paper has been performed with TOPCAT \citep{Taylor05}.

\end{document}